\documentclass[10pt,journal,compsoc]{IEEEtran}
\ifCLASSOPTIONcompsoc
  \usepackage[nocompress]{cite}
\else
  \usepackage{cite}
\fi
\ifCLASSINFOpdf
\else
\fi
\usepackage{amsmath,amsfonts}
\usepackage{array}
\usepackage{textcomp}
\usepackage{url}
\usepackage{graphicx}
\usepackage{cite}
\usepackage{subfig}
\usepackage{multirow}
\usepackage{xcolor}
\usepackage{mathrsfs}

\usepackage{colortbl}  

\usepackage{enumitem}
\usepackage{pifont}
\newcommand{\cmark}{\textcolor{green!80!black}{\ding{51}}}
\newcommand{\xmark}{\textcolor{red}{\ding{55}}}
\usepackage{amsthm}
\usepackage{thmtools} 
\theoremstyle{plain}

\theoremstyle{definition}

\theoremstyle{remark}

\usepackage[ruled,vlined,linesnumbered]{algorithm2e} 
\makeatletter
\newcommand{\nosemic}{\renewcommand{\@endalgocfline}{\relax}}
\newcommand{\dosemic}{\renewcommand{\@endalgocfline}{\algocf@endline}}
\let\oldnl\nl
\newcommand{\nonl}{\renewcommand{\nl}{\let\nl\oldnl}}
\makeatother

\begin{document}
%
\title{Rethinking Knowledge Distillation in Collaborative Machine Learning: Memory, Knowledge, and Their Interactions}
%
%
%
%

\author{Pengchao Han, Xi Huang, Yi Fang, and Guojun Han
\IEEEcompsocitemizethanks{\IEEEcompsocthanksitem Pengchao Han, Yi Fang, and Guojun Han are with the School of Information Engineering, Guangdong University of Technology, Guangzhou 510006, China. 
\protect\\
E-mails:  \{hanpengchao, fangyi, gjhan\}@gdut.edu.cn.
\IEEEcompsocthanksitem Xi Huang (corresponding author) is with Around Tech Company Ltd., Shenzhen, and also with Shenzhen Institute of Artificial Intelligence and Robotics for Society, Shenzhen 518000, China.  \protect\\E-mail:  huangxi@around-tech.com}
\thanks{This work was supported in part by the National Natural Science Foundation of China  (Grants 62401161, 62301336, 62322106, and 62471151),  Guangdong Basic and Applied Basic Research Foundation (Grant 2022A1515110056), the Science and Technology Program of Guangzhou (Grant 2025A04J3846), Longgang District Shenzhen's Ten Action Plan for Supporting Innovation Projects (project number: LGKCSDPT2024002).
}
\thanks{Manuscript received April 19, 2005; revised August 26, 2015.}}

\IEEEtitleabstractindextext{%
\begin{abstract}
Collaborative learning has emerged as a key paradigm in large-scale intelligent systems, enabling distributed agents to cooperatively train their models while addressing their privacy concerns. Central to this paradigm is knowledge distillation (KD), a technique that facilitates efficient knowledge transfer among agents. However, the underlying mechanisms by which KD leverages memory and knowledge across agents remain underexplored.
This paper aims to bridge this gap by offering a comprehensive review of KD in collaborative learning, with a focus on the roles of memory and knowledge. We define and categorize memory and knowledge within the KD process and explore their interrelationships, providing a clear understanding of how knowledge is extracted, stored, and shared in collaborative settings. We examine various collaborative learning patterns, including distributed, hierarchical, and decentralized structures, and provide insights into how memory and knowledge dynamics shape the effectiveness of KD in collaborative learning. Particularly, we emphasize task heterogeneity in distributed learning pattern covering federated learning (FL), multi-agent domain adaptation (MADA), federated multi-modal learning (FML), federated continual learning (FCL), federated multi-task learning (FMTL), and federated graph knowledge embedding (FKGE). Additionally, we highlight model heterogeneity, data heterogeneity, resource heterogeneity, and privacy concerns of these tasks. Our analysis categorizes existing work based on how they handle memory and knowledge. Finally, we discuss existing challenges and propose future directions for advancing KD techniques in the context of collaborative learning.
\end{abstract}

}

\maketitle

\IEEEdisplaynontitleabstractindextext

%
\IEEEpeerreviewmaketitle

\section{Introduction}\label{sec:intro}

\IEEEPARstart{R}{ecent} breakthroughs in deep learning, particularly in foundational models, have led to profound transformations across various sectors, including transportation, healthcare, and education. For instance, large language models (LLMs) like GPT-4 \cite{achiam2023gpt} and BERT \cite{koroteev2021bert} have demonstrated exceptional in-context learning capabilities in natural language understanding and generation across consecutive contexts, enabling a wide range of applications from conversational AI to advanced assistants for specialized domain tasks. These advancements prompt a re-examination of the similarities between foundational AI models and human cognition, particularly regarding the relationship between memory and learning, leading to the development of foundational model-based learning agents with increasingly sophisticated behaviors. 

These advancements also necessitate a rethinking of how learning agents will interplay in large-scale intelligent systems. In such systems, collaborative learning has emerged as a design paradigm for distributed intelligence. This approach allows various components of the system — each functioning as a learning agent — to cooperatively improve their models while addressing critical challenges in model heterogeneity, data heterogeneity, resource disparity, and privacy concerns. 

In collaborative learning, agents can be organized into various topologies, including distributed, hierarchical, and decentralized collaboration patterns. These configurations support a wide range of tasks, such as federated learning (FL) \cite{mcmahan2017communication}, multi-agent domain adaptation (MADA) \cite{farahani2021brief}, federated multi-modal learning (FML) \cite{lin2023federated}, federated continual learning (FCL) \cite{dong2022federated}, federated multi-task learning (FMTL) \cite{NIPS2017_6211080f}, and federated graph knowledge embedding (FKGE) \cite{chen2021fede}, etc. 
Effective collaborative learning relies on the sharing and reuse of learned knowledge across models, tasks, and agents to improve overall efficiency and performance. 
The typical knowledge transfer technique, i.e., knowledge distillation (KD) \cite{hinton2015distillingknowledgeneuralnetwork}, bridges the gap between heterogeneous agents by aligning representations and ensuring effective collaboration despite variations in data distributions or model architectures. 
Extensive research has leveraged KD to enable effective and efficient collaborative learning, addressing key challenges such as reducing communication overhead among agents, enhancing model architecture privacy, mitigating the forgetting of agents with respect to historical or external knowledge, and improving scalability across diverse participants.

Despite the extensive application of KD in collaborative learning, relatively little attention has been given to understanding the underlying rationale behind its effectiveness. Specifically, the mechanisms by which knowledge is stored, represented, and extracted during the KD process remain largely unexplored. The memory dynamics of a model and the nuanced process of transferring meaningful representations to other models are often overlooked, leaving gaps in understanding how KD facilitates learning in collaborative settings. 
The notions of \textit{memory} and \textit{knowledge} are widely employed in computer science \cite{sebastian2020memory,sun2023full,lu2024high,knuth1970neumann}, 
cognitive science \cite{squire2011cognitive,zacks2020event,nozari2024working}, neuroscience \cite{serences2016neural,josselyn2020memory,mankin2020modulation,sun2023organizing}, and artificial intelligence \cite{bietti2024birth,kim2024transformer,yao2024knowledge,ji2024linking,zhong2024memorybank,wang2024augmenting,yang2024text,park2023towards,allen2023towards,chen2021distilling,wang2021distilling,hinton2015distillingknowledgeneuralnetwork,gu2023remembering}. 
Their relationships and the boundaries of their contexts are mixed with vague or ambiguous interpretations.

By specifying the types of memory and knowledge utilized during collaborative learning, agents can gain deeper insights into underlying memory mechanisms and better assess the level of privacy protection within the collaborative framework. Locally encoded memory, such as model parameters, allows agents to extract knowledge more efficiently, streamlining the learning process. In practice, agents can store knowledge in encoded memory, leveraging caching techniques to enable rapid memory consolidation and facilitate swift knowledge extraction during collaborative interactions. Furthermore, investigating memory and knowledge in KD for collaborative learning provides a clearer understanding of the KD process. This, in turn, enhances the interpretability of knowledge transfer, offering a more transparent and systematic explanation of how knowledge is shared and utilized among agents.

Existing survey papers provide valuable insights into the diverse applications and methodologies of knowledge transfer and distillation in collaborative learning. For instance, Mora et al. \cite{mora2022knowledge} reviewed KD-based algorithms designed to address specific challenges in FL, such as model and data heterogeneity. Wu et al. \cite{wu2024knowledge} explored the role of KD in FL for edge computing, focusing on solutions for limited and heterogeneous resources, non-ideal communication channels, and the personalization of edge devices. Acharya et al. \cite{acharya2024survey} examined symbolic knowledge transfer from LLMs, highlighting methods to distill knowledge into interpretable symbolic representations, such as decision trees, logical rule sets, graphical models, or other structured forms that mimic neural network behavior. Furthermore, Liu et al. \cite{liu2024crowdtransfer} reviewed crowd knowledge transfer in Artificial Intelligence of Things (AIoT) communities, covering intra-agent and inter-agent (both decentralized and centralized) knowledge transfer. Their work emphasized knowledge transfer techniques, including domain adaptation, domain generalization, multi-task learning, KD, and meta-learning, with applications in human activity recognition, connected vehicles, and multi-robot systems.

Our paper distinguishes itself from the aforementioned work by providing a comprehensive review of KD in collaborative learning, analyzed through the lenses of memory and knowledge. The key contributions of our study are as follows:
\begin{itemize}
    \item \textbf{Definition of memory and knowledge, along with their relationships in KD}:
We define and categorize memory and knowledge within the context of KD and explore their interconnections, laying a solid foundation for understanding their roles in collaborative learning. 
    \item \textbf{Comprehensive overview of KD across various collaborative learning patterns}:
Our review encompasses how KD applies in collaborative learning paradigms, including distributed, hierarchical, and decentralized patterns, highlighting their unique characteristics and use cases.
    \item \textbf{Heterogeneity and privacy in distributed collaborative patterns}:
We focus on distributed learning, emphasizing task heterogeneity in scenarios such as  FL, MADA, FML, FCL, FMTL, and FKGE. Additionally, we examine model heterogeneity, data heterogeneity, resource heterogeneity, and privacy concerns of these tasks. Our analysis categorizes existing work based on how they handle memory and knowledge.
    \item \textbf{Challenges and future directions}:
We identify critical challenges in KD-driven collaborative learning and propose promising directions for future research to address these limitations.
\end{itemize}

\begin{figure*}[t]
	\centering \includegraphics[width=1.99\columnwidth]{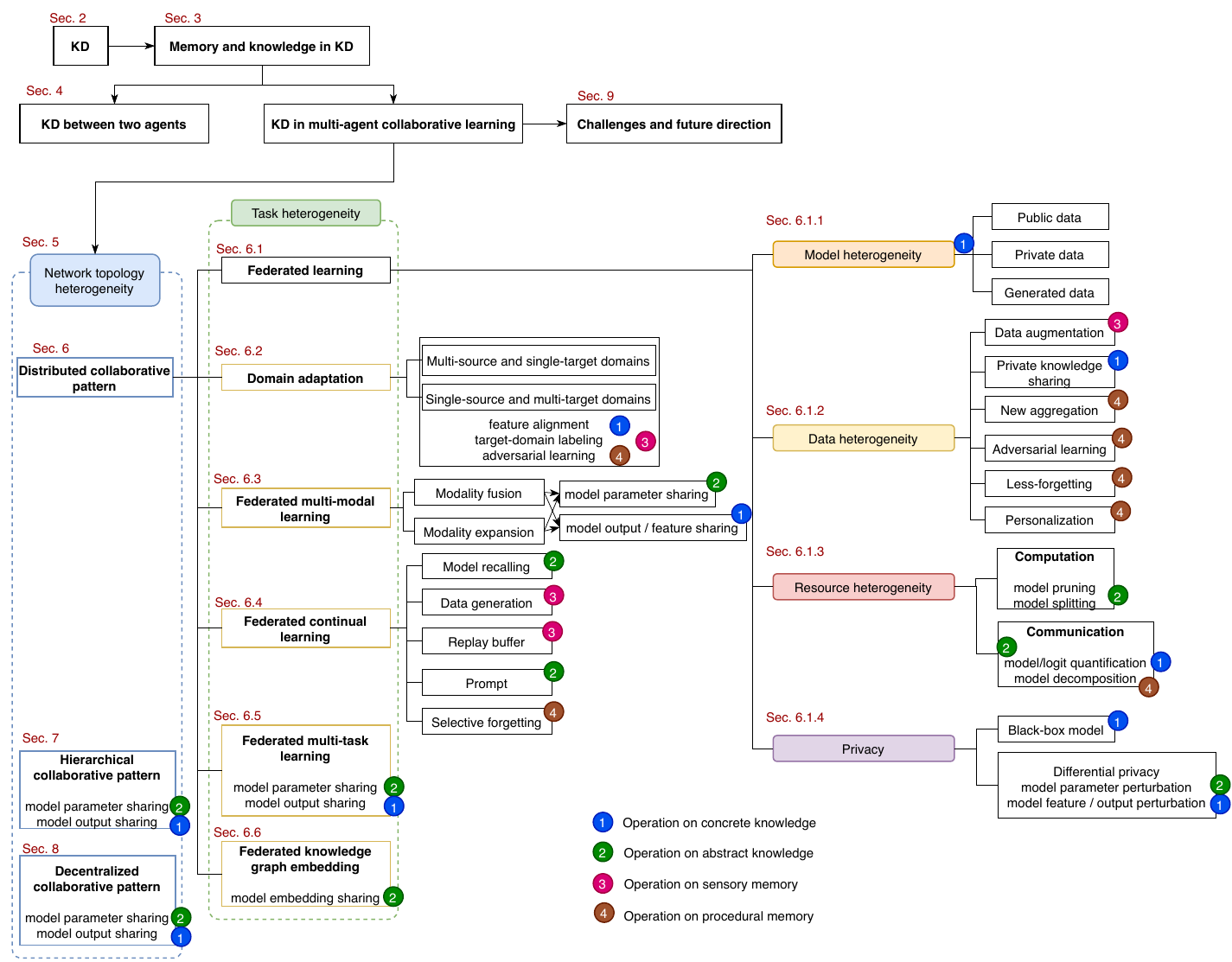}
	\caption{Organization of this paper.}
	\label{fig:content}
\end{figure*}

As illustrated in Fig. \ref{fig:content}, the rest of this paper is organized as follows. 
In Section 2, we review the development and workflow of KD, providing an overview of its evolution, rationality, variations, and application for collaborative learning. Section 3 delves into KD from the perspective of memory and knowledge, categorizing the types of memory and knowledge involved throughout the KD process. In Section 4, we analyze the memory and knowledge mechanisms underpinning KD interactions between two agents. Section 5 explores various collaborative patterns among multiple agents in the context of KD, establishing a foundation for subsequent discussions. Sections 6 through 8 review existing work on KD tailored for different collaborative patterns, highlighting its knowledge exchange in each context. Finally, Section 9 discusses the challenges and future directions in KD for collaborative learning, while Section 10 concludes the paper.

\section{Preliminary of Knowledge Distillation}
In this section, we take a brief review of the development of knowledge distillation in the past decade. Then, we demonstrate the typical workflow of KD as the preliminary to readers. 



\subsection{Development of Knowledge Distillation}
Knowledge distillation can be traced back to the seminal work \cite{hinton2015distillingknowledgeneuralnetwork} by Geoffrey Hinton and his co-authors. The idea of KD bears similarity with the work of Ba and Caruana \cite{ba2014deep}, in which they aimed to construct a lightweight neural network from a large-scale, complex model while maintaining comparable performance. Prior to that, the research community generally took a \textit{medium-inseparable} perspective of knowledge. That is, knowledge is \textit{not} separable from the memory medium. For example, knowledge of a neural network is implicitly encoded in the network's learned parameters. To transfer knowledge between models, early works \cite{domingos1997knowledge}\cite{buciluǎ2006model} employed data synthesized or labeled by learned models as the medium of knowledge between models. 
Such approaches rely on large amounts of data to encode and share learned knowledge inefficiently. When knowledge is transferred by utilizing a model to label data, 
the process ignores much information in the process of the model's inference. The issues become apparent in the cases where learning agents are not able to share the source data of knowledge in communication-limited regimes \cite{wang2023dafkd}\cite{han2024fedal}, where it is costly to train and tune a learning model from scratch \cite{aslam2023privileged}\cite{rao2023parameter}, and where learned models have numerous parameters that capture sufficient knowledge but perform cumbersomely when delivering responsive services (\textit{e.g.}, in recommendation systems \cite{jin2023feddyn}\cite{li2024contextual}).

Instead, canonical KD methods \cite{gou2021knowledge, wang2021knowledge,li2022distilling} take a different perspective by considering knowledge representation \textit{separable} from the underlying network structure. 
They view knowledge in a neural network as an abstract mapping from the input to the output for a particular task. An instance of such mapping consists of a concrete input and some output from the network extracted through some distillation mechanism. 
Unlike the approaches based on the medium-inseparable perspective, KD utilizes much richer information in the inference process to elicit generalized knowledge about the input (\textit{e.g.}, class logits when provided with a picture of particular patterns in a visual classification task \cite{hinton2015distillingknowledgeneuralnetwork}). 
Such a perspective has motivated continuous development of KD approaches in the past decade, such as logit-based, feature-based, and relation-based KD mechanisms \cite{gou2021knowledge}\cite{ liu2021data}.

Parallel to the development of KD, recent years have witnessed the emergence of foundation models \cite{bommasani2021opportunities}. Such models are considered pivotal for advancing artificial general intelligence applications as they demonstrate remarkable performances across various task domains, largely driven by neural scaling laws \cite{alabdulmohsin2022revisiting,bahri2024explaining,michaud2024quantization}.  
Several recent studies \cite{jiang2024large,xu2024large,shen2024large,huang2024toward} investigated the potential for deploying foundation models in network edge environments to deliver responsive services. However, this transition encounters several concerns, including 1) the mismatch between substantial computing resource and data demands of foundation models and the resource-constrained nature of network edges \cite{shen2024large}, 2) the capability and scalability of managing knowledge in smaller models \cite{ojha2023knowledge}\cite{schaeffer2024emergent}, and 3) the explainability of transferred knowledge \cite{acharya2024survey}.

Recent research addresses resource concerns by introducing \textit{scale-down} methods, which compress large foundation models by distilling their knowledge into smaller models. Empirical studies \cite{fu2023specializing,wang2024can,ko2024distillm,hsieh2023distilling,wu2024divide,tian2024tinyllm} have shown that these scale-down techniques can significantly enhance the resource efficiency of model deployment. However, a common drawback of these methods is that the limitations of smaller models often lead to a decline in overall performance or task-specific effectiveness. 

To address concerns about capability preservation, another avenue of recent research embraces a \textit{scale-out} paradigm. In this approach, the design of large models is conceptualized as modular circuits \cite{weiss2022neural}\cite{ge2024circuits}, wherein knowledge tailored to specific tasks is empirically found concentrated in some subsets of model parameters or neural modules \cite{dai2022knowledge,guan2024knowledge,niu2024does}. This perspective not only facilitates the analysis of complex behaviors manifested by large foundation models but also encourages the decomposition of large models into smaller, editable components \cite{merullo2024circuit}\cite{de2021editing}. Motivated by this perspective \cite{yao2024knowledge}\cite{yang2024text}, some recent works distinguish \textit{local} knowledge from \textit{global} knowledge in a model. Local knowledge refers to the mapping of specific instances to expected outputs, which is localized within a critical subgraph of the model. This knowledge can be separated from the original model, stored in external memory medium \cite{zhong2024memorybank}\cite{yang2024text}, and retrieved for use in relevant contexts. In contrast, global knowledge encompasses all remaining information that is not distinctly contained within any subgraph of the model. Distinguishing between local and global knowledge allows for the structural distillation of knowledge from large models. This separation facilitates the reduction of deployed model sizes by managing local knowledge in external storage while also enabling flexible sharing of knowledge among different learning agents. 

While the scale-down and scale-out approaches address resource and scalability concerns, the distillation process remains opaque and lacks interpretability \cite{acharya2024survey}. To tackle this issue, several studies have explored converting distilled knowledge into symbolic forms \cite{hu2016harnessing,chen2022probing,gupta2023neurosymbolic,aakur2023leveraging}. This conversion enhances the interpretability of the distillation process, improves sample efficiency, and facilitates the integration of human guidance or annotated data.

\subsection{Workflow and Applications of KD}
The KD process within the teacher-student framework typically involves two stages:
\begin{itemize}
    \item \textbf{Training Stage}: In this stage, large and complex models (such as deep neural networks) are trained on a comprehensive dataset, achieving high performance by effectively capturing intricate patterns and relationships within the data. Each model, known as a \textit{teacher} model, serves as a benchmark for knowledge and competency in a specified task. 
    \item \textbf{Distillation Stage}: During this stage, a smaller, more streamlined model — referred to as the \textit{student} — is trained to imitate the soft (intermediate) output of the teacher model. 
    Specifically, the soft outputs, such as logits or the probability distribution generated by the teacher for various outputs, contain much richer information than the teacher's direct outputs. By optimizing its own parameters, the student model aligns its predictions with these soft outputs and, when available, the reference outputs or ground truths. This approach enhances the student’s ability to capture the underlying patterns in the data, improving its overall performance. 
\end{itemize}
In the second stage, a common loss function is the Kullback-Leibler (KL) divergence, which measures the difference between the teacher’s and student’s soft logits \cite{zhang2018deep,shejwalkar2019reconciling}. Another common choice is the cross-entropy loss \cite{Distilling,anil2018large}. 

\textbf{Main advantages.} KD offers several key benefits:
\begin{itemize}
    \item Model compression: KD enables the training of a smaller student network from a large pre-trained teacher network, effectively reducing model size without significant loss of performance \cite{ashok2017n2n}. 
    \item Generalization improvement: The student model often generalizes better by learning from the teacher’s soft targets, which provide richer information than one-hot labels. Additionally, training multiple networks with the same architecture but different initializations can improve diversity; KD can then be applied from better-performing networks to weaker ones to enhance overall performance \cite{hong2021periodic}.
    \item Knowledge transfer: KD facilitates the transfer of knowledge from a powerful model trained on a large dataset to a smaller model, making it particularly useful for tasks with limited data or computing resources \cite{lin2017collaborative}. 
\end{itemize}

\textbf{Rationality behind KD.} Researchers have explored the underlying rationale for KD from multiple perspectives:

First, from the point of KD loss, KD can be formulated as a maximization of mutual information between the representations of the teacher and student networks \cite{tian2019contrastive,Variationalinformation,wang2021knowledge}. By analyzing the gradient of the KD loss, it has been observed that the distilled loss effectively aligns the logits between the teacher and student models \cite{gou2021knowledge}. In terms of gradient propagation, KD can be regarded as an adaptive version of label smoothing, which implies that it inherits many regularization benefits from label smoothing, such as improved model regularization, better calibration, and reduced overconfidence \cite{tang2020understanding}.  Furthermore, \cite{zhao2022decoupled} decouples the KD loss into components for the target class and non-target classes. By balancing these components, KD can be applied more efficiently and flexibly.

Empirical evidence from \cite{cheng2020explaining} suggests that KD helps models learn more visual concepts than they would from raw data alone. Moreover, KD encourages deep neural networks (DNNs) to learn various visual concepts simultaneously, as opposed to the sequential learning observed when training directly from raw data. Additionally, KD provides more stable optimization directions than direct learning from raw data.

Theoretically, the effects of KD can be understood through VC theory \cite{vapnik1999nature,lopez2015unifying,mirzadeh2020improved,menon2020distillation} and domain adaptation \cite{lin2020ensemble}, which suggest that learning from a teacher model reduces the search space of functions, leading to more efficient learning compared to training from raw data.

\textbf{Variations for performance improvement.} While KD traditionally relies on soft logits (i.e., the model's output) for knowledge transfer, various approaches have been proposed to enhance its performance. These include adjusting the temperature parameter to a higher value \cite{shejwalkar2019reconciling}, incorporating feature distillation \cite{romero2014fitnets}, and utilizing online distillation \cite{9892595} and co-distillation \cite{anil2018large}.

In KD, the temperature is a hyperparameter that controls the ``softness'' of the probability distribution generated by the teacher model. Specifically, the softmax function that produces output probabilities from the logits is adjusted by dividing the logits by the temperature. When a higher temperature is used, it results in smoother probabilities, allowing the student model to learn from the teacher in a more nuanced way. This means the student doesn’t just focus on the most probable class but also gains insights from the similarity among classes, which can improve generalization. Additionally, a higher temperature can help prevent the student model from overfitting to the teacher’s precise outputs, particularly if the teacher model is over-parameterized. The softer distribution ensures that the student does not learn overly specific patterns that might not generalize well to unseen data.
However, while high temperatures provide richer and smoother learning, setting the temperature too high might cause the student model to focus excessively on irrelevant classes, potentially diluting the meaningful distinctions between the correct class and other classes. Conversely, a lower temperature sharpens the teacher's output probabilities, making them more similar to hard labels. This approach can be beneficial when the student model needs to concentrate more on accurately identifying the correct class rather than learning from soft probabilities. Hence, it is crucial to find appropriate temperatures to achieve efficient KD.

Feature distillation is a variant of KD where, rather than focusing on the final output (logits or softmax probabilities), the goal is to transfer knowledge from the internal representations or feature maps of the teacher model to the student model \cite{romero2014fitnets, zagoruyko2016paying, park2019relational}. In \cite{NEURIPS2023_cdddf13f}, the authors enhance this approach by extracting denoised features from both the teacher and student models using a Diffusion model, which supports the distillation process \cite{ho2020denoising}. These internal representations capture the hierarchical features learned by the teacher model during training. Compared to traditional KD, feature distillation allows for a richer knowledge transfer, often leading to improved model performance. A comprehensive survey on feature distillation can be found in \cite{heo2019comprehensive}.

Beyond vanilla KD, online distillation methods, such as \texttt{FLM} \cite{9892595}, have been proposed to address the challenge of an unavailable teacher model. \texttt{FLM} trains multiple models with heterogeneous architectures simultaneously, fostering diversity among them. This approach leads to improved model performance compared to vanilla KD by leveraging the combined output of these diverse models.

In systems with multiple coexisting agents, co-distillation \cite{anil2018large} is often employed to enable mutual knowledge transfer among agents. The average of agents' model output or encoded data features has been utilized in various federated learning approaches. For instance, federated knowledge distillation \cite{seo2020federated} facilitates collaboration among multiple agents through online co-distillation.

Applications of KD for collaborative learning. 
Beyond model compression, KD has proven valuable in facilitating knowledge transfer among multiple agents for diverse purposes. These include  federated learning \cite{li2019fedmd}, domain adaptation \cite{zhao2020multi}, federated multi-modal learning \cite{9567675}, federated continual learning \cite{ma2022continual}, and federated multi-task learning \cite{Kundu_2019_ICCV}. Additionally, 
KD has been effectively applied to a wide range of scenarios for various objectives. 
\begin{itemize}
    \item \textbf{Model Compression}: KD has been instrumental in compressing large models for efficient deployment without significantly sacrificing performance. Applications include medical melanoma detection \cite{khan2022knowledge}, object detection \cite{yadikar2023review, zhang2023structured, wang2024crosskd}, semantic segmentation \cite{ji2022structural}—which involves assigning a unique category label to each pixel in an input image—speech recognition \cite{lu2017knowledge}, face recognition \cite{li2023rethinking, ge2018low}, image super-resolution \cite{zhang2021data}, and image-to-image translation \cite{zhang2022wavelet}.
    \item  \textbf{Transfer Learning}: KD enables knowledge transfer from a rich, high-capacity teacher model—often trained on multiple tasks or modalities—to a lightweight student model. This is particularly effective when leveraging the teacher’s ability to extract high-level, transferable features from complex datasets.
For example, distilling from multi-task models to a single task of natural language understanding \cite{liu2019improving}, transferring knowledge from models trained on multi-modal data to single modality for applications include recommendation systems \cite{liu2023semantic, kang2020rrd}, action recognition \cite{thoker2019cross}, medical image segmentation \cite{wang2023prototype}, MRI segmentation \cite{9567675}, and sentiment analysis \cite{li2024correlation}.

\item \textbf{Data-Scarce Scenarios}: KD is particularly valuable when labeled data is limited. By leveraging the rich supervision from a teacher model, KD can help generalize better in low-data regimes. 
Open-Vocabulary Object Detection for detecting objects based on arbitrary textual descriptions \cite{gu2021open}, distilling knowledge from LiDAR-enhanced models to improve radar-based object detection \cite{bang2024radardistill}, transferring of action unit knowledge to recognize subtle facial expressions \cite{sun2020dynamic}, trajectory forecasting \cite{monti2022many}, and video captioning \cite{pan2020spatio}.
\end{itemize}

\section{Rethinking KD from lens of Memory \& Knowledge}
\begin{figure}[t]
	\centering \includegraphics[width=0.7\columnwidth]{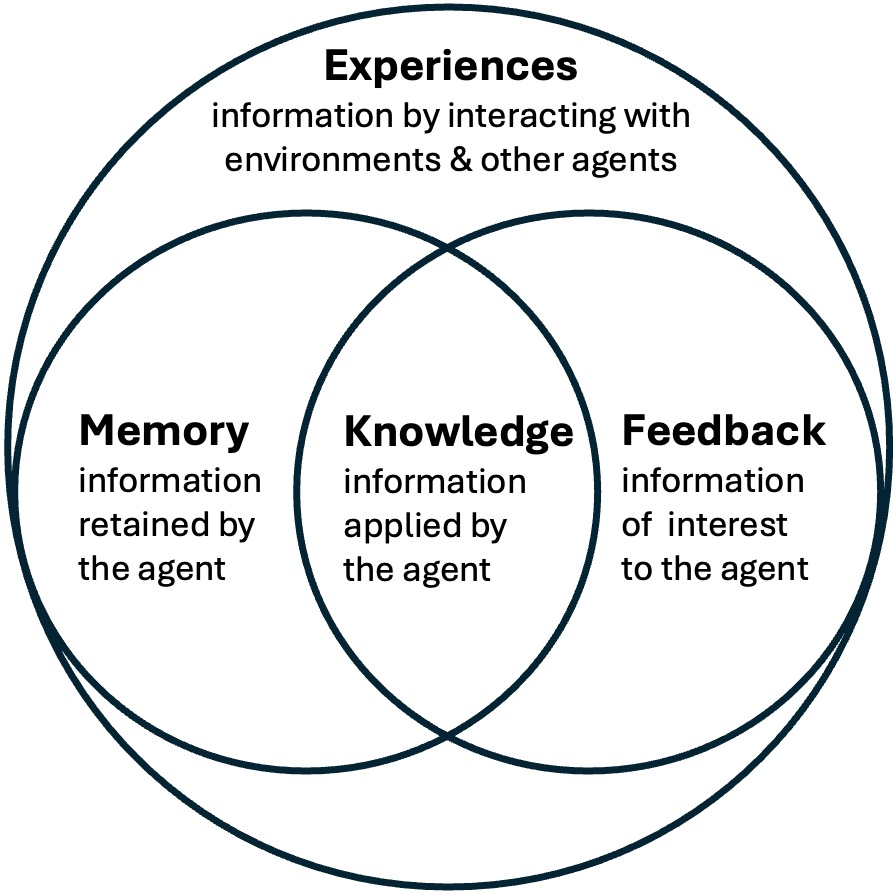}
	\caption{Relationship between knowledge, memory, and experiences of agent.}
	\label{fig:mem}
\end{figure}

\begin{figure*}[t]
	\centering \includegraphics[width=1.6\columnwidth]{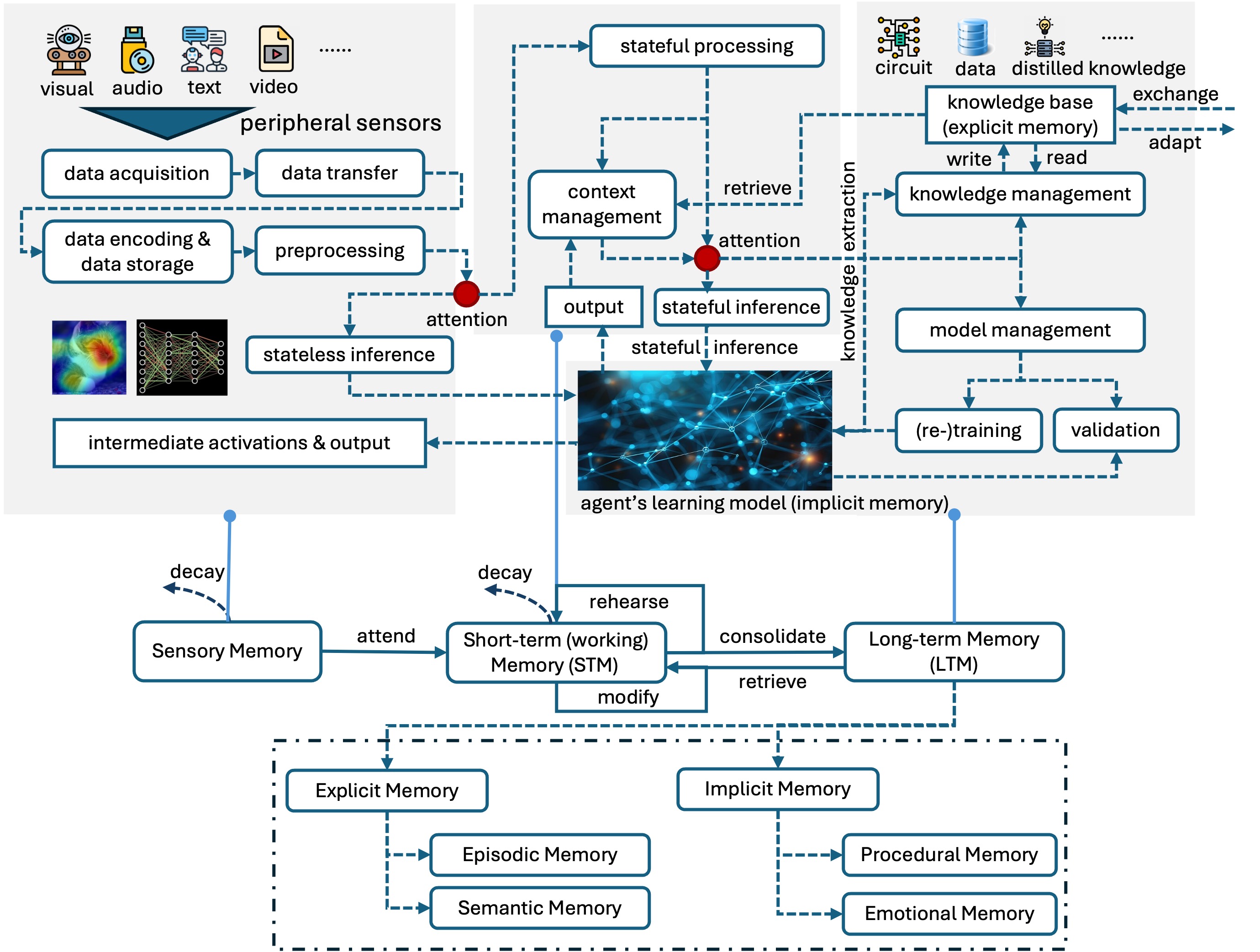}
	\caption{An illustration of memory hierarchy and its relation with agent-based learning.}
	\label{fig:mem-1}
\end{figure*}


While most existing works regard KD as a process for transferring knowledge from larger models to smaller ones, they often overlook the subtle distinction between knowledge and memory. In modern AI, particularly within deep neural networks (DNNs), knowledge refers to the hidden, compressed representations that encode patterns and relationships derived from training data. This knowledge is captured in various forms, depending on the structure of the underlying medium and the mechanisms used to extract and encode that knowledge - commonly referred to as memory mechanism. For a single DNN, the memory mechanism encompasses both its network architecture and the training approach. In typical KD, the memory mechanism includes the neural network structures of both the teacher and the student models, as well as their respective training schemes at each stage. 

Achieving decoupling between knowledge and memory in knowledge distillation involves several strategies. One approach is to design models that can separate the representation of knowledge from the mechanisms that store it. This could be accomplished by employing modular architectures, where knowledge is represented in distinct layers or network components that can be independently accessed and manipulated.

Another strategy is to utilize transfer learning techniques that allow the student model to adapt and refine its knowledge without being overly reliant on the specific memory structures of the teacher model. By being trained on diverse datasets or tasks, the student will develop its representations of knowledge, independent of the teacher's memory.

Additionally, incorporating techniques such as attention mechanisms can help the student model focus on relevant information while disregarding the specific memory patterns of the teacher. This allows for a more flexible interpretation of knowledge, enabling the student to generalize better across different contexts.

Finally, exploring alternative forms of knowledge representation, such as symbolic or probabilistic methods, may provide insights into how knowledge can be effectively decoupled from memory. By representing knowledge in a way that is less tied to the underlying architecture, we can foster a more robust understanding that transcends specific memory implementations.

In the context of knowledge distillation, the teacher model, having accumulated knowledge from extensive training, serves as a rich source of memory that encapsulates complex patterns and relationships within the data. This model not only provides hard outputs but also conveys soft outputs—probability distributions that encode nuanced insights about data representation. The student model, in turn, acts as a learner that leverages this interwoven web of knowledge. By absorbing the soft outputs from the teacher, the student can effectively compress and retain this rich information, transforming it into a more manageable form of memory. This process allows the student to build an efficient representation of knowledge that can influence its decision-making and enhance its performance in subsequent tasks.

Furthermore, the process of aligning the student model’s learning with the teacher's encoded knowledge can be viewed as an intricate mechanism of memory consolidation. As the student model receives both the soft targets from the teacher and the reference outputs, it engages in an active process of memory formation, where critical insights are extracted and stored. This dual approach not only optimizes the learning efficiency of the student but also allows it to establish meaningful associations between features of the input data and the teacher's learned representations. Consequently, knowledge distillation can be understood as a phenomenon that not only facilitates the transfer of knowledge but also embeds a form of memory architecture within the student, enabling it to adapt and respond intelligently to novel situations. This perspective highlights the richness of knowledge distillation as a learning paradigm that emphasizes the critical roles of memory and knowledge in shaping the capabilities of AI agents.

As illustrated in Figure \ref{fig:mem}, both \textit{memory} and \textit{knowledge} are fundamental components of learning agents. However, their characteristics are highly dependent on the design of the agents, reflected by the heterogeneity in their underlying neural network design. This variability requires training each agent from scratch in a less cost-effective fashion. To mitigate this issue, it is beneficial to reuse and transfer knowledge from well-trained learning agents to those that are not. This idea has led to the inception of knowledge distillation \cite{hinton2015distillingknowledgeneuralnetwork}. 

In this section, we first review the basics of knowledge distillation. Then, we switch to the development 
of memory and knowledge from the lens of KD, followed by a presentation of memory and knowledge involved in this paper.

\subsection{Rethinking Memory in Knowledge Distillation}

Memory mechanisms have long been fundamental yet often overlooked in machine learning. As a pivotal branch of this field, deep learning highlights the importance of memory in modeling and processing information, enabling systems to learn from past experiences and adapt to new situations. Inspired by the information-processing cognitive model proposed by Atkinson and Shiffrin \cite{shiffrin1969storage}, many deep learning techniques are rooted in the neural connections present in the human brain. Consequently, memory is typically thought to be encoded within the connections of artificial neurons in these networks. This memory-centric design has formed the foundation for modern AI and has demonstrated remarkable performance in real-world applications.

In multi-agent environments, the integration of effective memory mechanisms can significantly enhance collaboration, coordination, and learning efficiency among agents. Agents are required not only to navigate their immediate surroundings but also to consider the actions and strategies of other agents. Memory mechanisms empower agents to store and retrieve relevant information from past interactions, which is crucial for understanding opponents' behaviors and predicting future actions. This capability can enhance decision-making strategies, particularly in dynamic, adversarial, or cooperative settings.

\subsection{Categorizing of Memory Mechanisms}

Inspired by the Atkinson-Shiffrin memory model, we classify the memory mechanisms in multi-agent systems as follows, illustrated by Figure \ref{fig:mem-1}. 

\begin{itemize}
    \item \textbf{Sensory memory} is the initial stage where sensory information from the environment is briefly held. It is characterized by its short duration (milliseconds to seconds) and large capacity. For example, in an agent-learning system, the peripheral sensors would be responsible for capturing raw sensory data, such as visual inputs from cameras or auditory inputs from microphones. This data is then processed by the data acquisition component, which further refines the sensory information before it is passed on to subsequent stages.
    \item \textbf{Short-term memory} (STM), also known as working memory, is a temporary storage system for holding a small amount of information for a short period (seconds to minutes). It is involved in active processing and manipulation of information. In an agent-learning system, the attention mechanism would select relevant information from the sensory input, focusing on specific features or patterns. This selected information is then processed by the stateless inference component, which performs initial computations and feature extraction. The stateful inference component, on the other hand, maintains a representation of the agent's current state and updates it based on the incoming information.
    \item \textbf{Long-term memory} (LTM) is a vast storehouse for information that can be retained for extended periods (hours, days, years). It is divided into two main types: explicit memory and implicit memory. In an agent-learning system, the knowledge base would store learned knowledge and experiences. The knowledge management component would be responsible for organizing and updating the knowledge base, while the model management component would oversee the training and adaptation of the agent's learning models. The rehearse component would facilitate the consolidation of information from short-term memory to long-term memory, ensuring that important experiences are retained for future use. 
    
    Two types of memory mechanisms are considered in explicit memory. 
    \begin{itemize}
        \item \textbf{Semantic memory} is a type of explicit memory that stores factual knowledge and concepts. It is organized hierarchically, allowing agents to relate different pieces of information and draw inferences. For example, an agent might learn the concept of ``cat" and associate it with other related concepts, such as ``feline," ``mammal," and ``pet." This semantic knowledge can be used to understand new information, make predictions, and solve problems.
        \item \textbf{Episodic memory} is another type of explicit memory that stores personal experiences and events. It is organized in a temporal sequence, allowing agents to remember the order in which events occurred and the context in which they happened. For example, an agent might remember a specific instance of encountering a cat, including the location, time, and behavior. Episodic memory is crucial for agents to learn from their experiences, adapt to changing environments, and make informed decisions. This might involve: 1) \textbf{Storing experiences} -- recording specific events or interactions that the agent has encountered, including the time, location, and relevant details; 2) \textbf{Retrieving experiences} -- recalling past events to inform future decisions or actions; and 3) \textbf{Updating experiences} -- modifying stored experiences based on new information or to reflect changes in the agent's understanding of the world.
    \end{itemize}
    Implicit memory can be divided into two primary types: procedural memory and emotional memory. 
    \begin{itemize}
        \item \textbf{Procedural memory} refers to the unconscious storage of skills and habits, enabling individuals to perform tasks without needing to consciously recall each step. For example, riding a bicycle or playing a musical instrument involves relying on procedural memory, where the brain has internalized the sequence of actions through repetition and practice.
        \item \textbf{Emotional memory} involves the automatic recall of emotional responses to past experiences. These memories are often linked to specific events that triggered strong feelings, such as fear, joy, or sadness. Emotional memory plays a crucial role in shaping behaviors and responses to similar situations in the future. For instance, someone who has experienced trauma may have an emotional memory that causes heightened anxiety in similar contexts, even without consciously remembering the original event.
    \end{itemize}
\end{itemize}

In multi-agent collaborative learning, agents encoded raw data into semantic memory or procedural memory through model training. 

\subsection{Rethinking Knowledge in Knowledge Distillation}
Based on the previous review, we categorize knowledge in KD from the following two orthogonal dimensions. 

\textbf{Dimension} 1: \textit{Medium-separability}. If knowledge representation is fully inseparable from its memory medium (\textit{e.g.}, neural network model or dataset), such knowledge is called \textit{medium-inseparable}. The learned parameters in a neural network and the sample-label pairs in a dataset are both examples of medium-inseparable knowledge. If the representation is not fully inseparable from its medium, then it is called \textit{medium-separable}. For example, in logit-based KD, the inputs and their corresponding logit outputs are medium-separable knowledge. Another example is in symbolic KD where the distilled knowledge is converted into interpretable symbols. In this case, the symbolic knowledge is medium-separable. 

\textbf{Dimension} 2: \textit{Knowledge-locality}. If a piece of knowledge is localized within a small part of its memory medium (\textit{e.g.}, neurons or neural modules when encoded in a neural network), it is called \textit{local} knowledge. Otherwise, it is called \textit{global} knowledge. For example, given concrete input, the subgraph in a neural network which encodes generalized knowledge about its output is local knowledge (\textit{a.k.a.} knowledge circuits in \cite{yang2024text}). The remaining components of the network, which do not localize knowledge specific to any particular input but still contribute to the inference, are taken as global knowledge. 

The medium-separability of global and local knowledge can vary significantly depending on the context.
In a neural network, global knowledge refers to the parameters of components that do not localize knowledge to any specific input, and as such, it remains medium-inseparable. Conversely, in text databases, global knowledge can encompass the overarching logical relationships among different entities within the corpus. This type of knowledge does not reside in any particular part of the dataset but can be represented in alternative formats, such as knowledge graphs. In this instance, global knowledge is medium-separable, allowing it to be extracted and utilized independently of the original data structure.

On the other hand, when local knowledge pertains solely to specific areas within a neural network, it remains medium-inseparable. However, if mechanisms are in place to separate these areas from the original network and store them in different memory media \cite{zhong2024memorybank}\cite{yang2024text}, then local knowledge becomes medium-separable. In this scenario, local knowledge is not only editable but can also be directly shared among different learning agents that adhere to the established mechanism. This capability allows for greater flexibility and collaboration among agents, as they can utilize and adapt each other's localized knowledge to enhance their learning processes.

\begin{figure}[t]
	\centering \includegraphics[width=0.9\columnwidth]{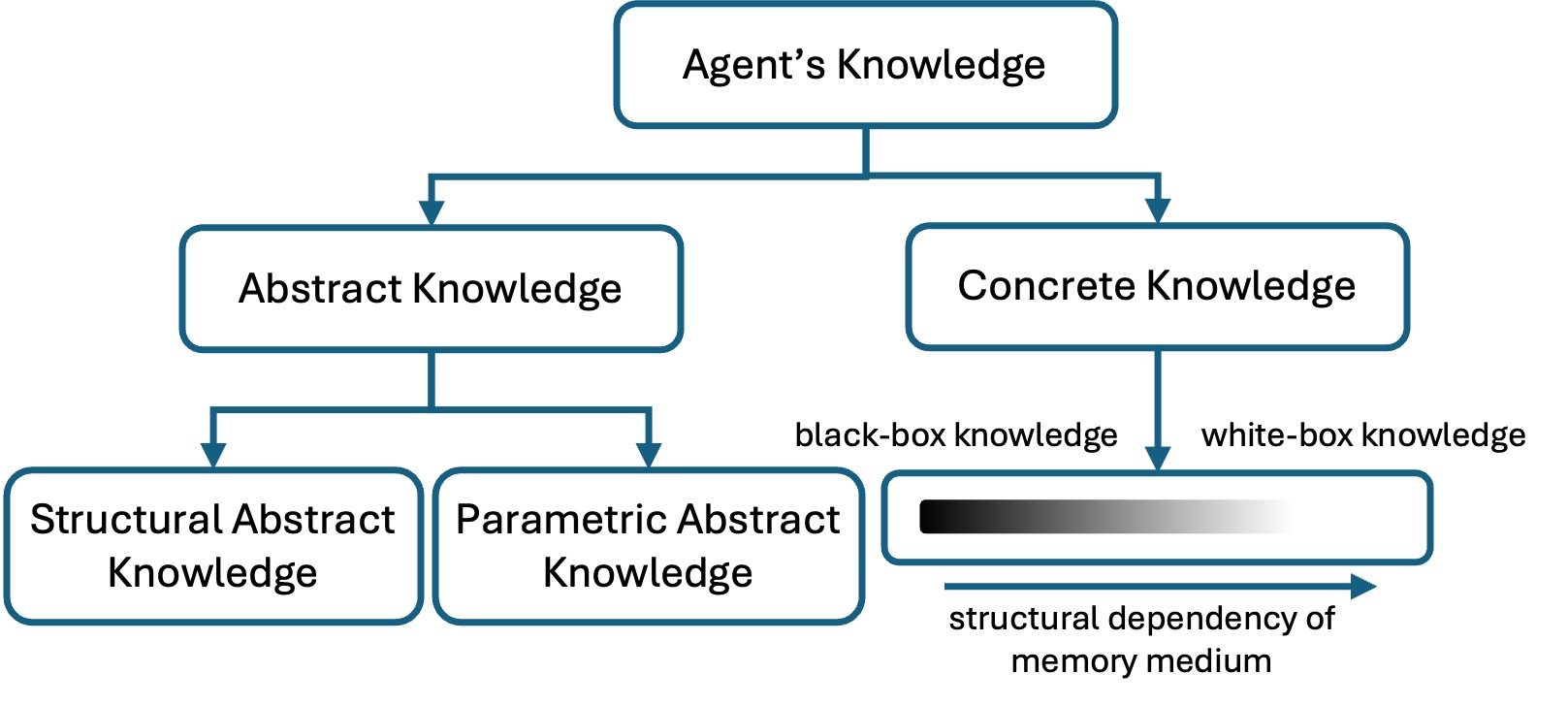}
	\caption{An illustration of knowledge for agent-based learning.}
	\label{fig:mem-2}
\end{figure}

\textbf{Dimension 3}:\textit{ Semantic property.}
Depending on whether the information in a knowledge corresponds to concrete or abstract entities, knowledge can be classified into \textit{concrete knowledge} and \textit{abstract knowledge}, as illustrated in Fig. \ref{fig:mem-2}. 
\begin{itemize}
    \item \textbf{Concrete knowledge} refers to knowledge that involves facts, events, and opinions about the world that can be clearly demonstrated. 
    
    In the above definition, facts are the known states of things and are presented as-is without the need for induction or deduction; events are the known changes in the states of things; opinions reflect individual beliefs about the states of things. These types of entities are organized in concrete knowledge and ready for later use in inference and reasoning. 

The concrete knowledge can be \textit{black-box} or \textit{white-box}, 
depending on whether or not it involves dependency on the underlying medium they are encoded upon. 
If the knowledge is independent of the medium, then it is black-box, and white-box otherwise. 
An example of black-box concrete knowledge is the form of knowledge extracted from the teacher in KD \cite{Distilling}, while an instance of white-box concrete knowledge has been described in the recently developed memory circuit theory \cite{yang2024text}.  
    \item \textbf{Abstract knowledge} refers to knowledge that organize a set of abstract concepts which interact with events, facts, and opinions non-trivially in a certain context. 

Abstract memory can be categorized into two types: structural and parametric. Structural abstract memory refers to the organization or architecture of the information, where the relationships between data points or concepts are preserved in a way that reflects their inherent structure. For instance, in knowledge graphs or neural networks, the connections between nodes or layers capture the underlying relationships, enabling the system to reason about the structure of the data.

On the other hand, parametric memory focuses on the storage of information in the form of parameters, which are adjusted based on the learning process. These parameters encode essential features of the data in a compressed or simplified format, allowing for efficient retrieval and decision-making. A notable example is the weight matrices in deep learning models, where the parameters learned during training encapsulate critical information about the task at hand.
\end{itemize}

In this paper, we focus on the semantic property of knowledge and its relation to memory.

    


    





    



\section{Rethinking KD Between Two Agents}

In this section, we begin by introducing the forms and extraction methods of knowledge in deep learning, covering both abstract and concrete knowledge. Next, we discuss the roles of memory and knowledge in facilitating KD between two agents.

\subsection{Forms and Extraction of Abstract Knowledge}
Achieving good performance in deep learning requires a large amount of raw data, which can compromise an agent's privacy or induce high communication overhead among multiple agents for collaborative learning. Agents can encode raw data, i.e., sensory memory, into abstract knowledge to facilitate preserving privacy and reducing communication overhead, enabling effective collaborative learning. Specifically in deep learning, the abstract knowledge takes the form of model parameter and knowledge graph: 
\begin{itemize}
    \item Model Parameter: The information from raw data can be encoded into neural network models. By sharing these model parameters, any agent can inherit the predictive capabilities of the agent that owns the raw data.
    \item Knowledge graph: Knowledge graph models the relationship among entities. A knowledge graph is a structured representation of knowledge, typically used to organize and interlink information in a way that can be easily understood and processed by both humans and machines. Knowledge graphs consist of nodes (representing entities) and edges (representing relationships between entities). Key characteristics of a knowledge graph include:
\begin{itemize}
    \item Entities: The nodes in a knowledge graph represent real-world objects, concepts, or people.
    \item Relationships: The edges represent the relationships between these entities (e.g.,``developed," ``is a type of").
    \item Attributes: Each entity or relationship can have attributes or properties that provide additional information (e.g., the birthdate of a person, the publication date of a book).
\end{itemize}
\end{itemize}

For abstract knowledge, model parameters and knowledge graphs can be obtained through model training and directly shared among agents. 

\subsection{Forms and Extraction of Concrete Knowledge}
Concrete knowledge refers to information or facts that are specific, tangible, and easily understood. It typically involves clear, detailed, and often observable data that can be directly experienced or measured. In deep learning, a typical form of concrete knowledge is model (intermediate) output that corresponds to a specific input:
\begin{itemize}
     \item Model (intermediate) Output: The input-output relationships can reflect the dark knowledge in the raw data. By using representative data and extracting the input-output relationships of the neural network model, an agent can learn the knowledge more easily.
\end{itemize}

For model intermediate features and outputs, it is crucial to properly select input data samples and derive input-output relationships from model parameters. Generally, there are the following methods to extract model outputs:
\begin{itemize}
    \item Using public dataset \cite{li2019fedmd}: When all agents share a set of public data, concrete knowledge can be extracted by computing the model outputs of these public data samples. Public datasets typically contain a small volume of unlabeled data, which is easier to obtain than labeled private data.
    \item Using private dataset \cite{jeong2018communication,park2019distilling}: In cases where a public dataset is unavailable, agents can extract concrete knowledge from their private datasets. To preserve privacy, only averaged intermediate features or model outputs computed from a batch of data are shared among agents.
    \item Using generated dataset \cite{zhang2021fedzkt,9879661}: Relying solely on private datasets and sharing average model outputs can limit the amount of knowledge extracted. An alternative approach is to generate a public dataset to facilitate the sharing of input-output relationships among agents. Advanced techniques such as Generative Adversarial Networks (GANs) and diffusion models can be used for this data generation.
    \item Prompt \cite{liu2023federated}: A prompt refers to the input text or question given to the model to generate a response. The prompt serves as a starting point for the model to produce a relevant and coherent output. The prompt is generally used in language models.
\end{itemize}

\subsection{Memory \& Knowledge in KD  Between Two Agents}

\begin{figure}[t]  
	\centering     
	\subfloat[Concrete knowledge sharing] 
	{
			\centering           
			\includegraphics[width=0.45\textwidth]{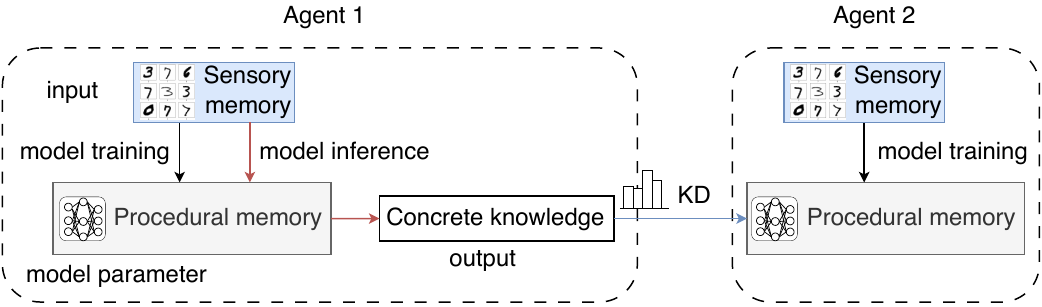}   \label{fig:kd2-a}
	}
 \hfill
	\subfloat[Abstract knowledge (model parameter) sharing]  
	{
			\centering       
			\includegraphics[width=0.5\textwidth]{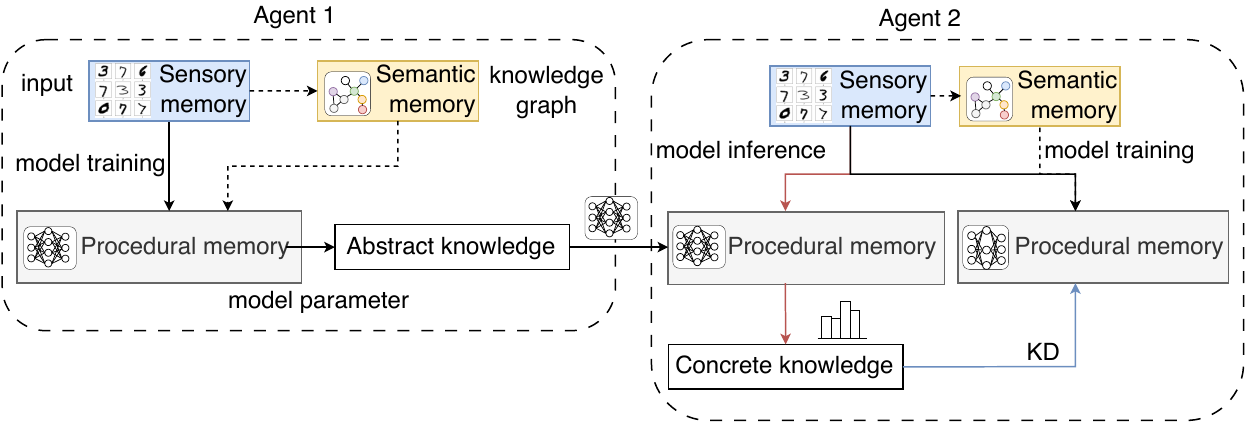}   \label{fig:kd2-b}
	}%
 \hfill
 \subfloat[Abstract knowledge (knowledge graph) sharing]  
	{
			\centering       
			\includegraphics[width=0.45\textwidth]{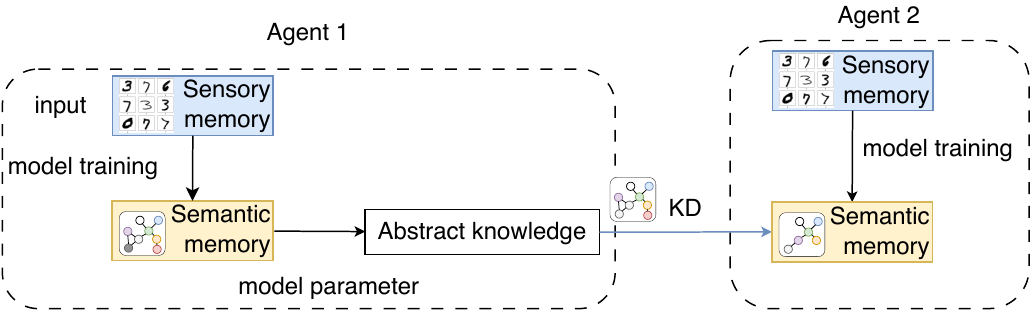}  \label{fig:kd2-c} 
	}%
	\caption{An illustration of memory and knowledge in KD between two agents. } \label{fig:kd-two}
\end{figure}
There are two primary types of knowledge that can be shared between agents in KD: concrete knowledge and abstract knowledge. 
In standard KD, a client shares its model outputs to facilitate knowledge transfer. The memory process for this type of KD is illustrated in Fig. \ref{fig:kd2-a} and proceeds as follows:

\begin{itemize} 
\item Agents first collect and store raw data as sensory memory; 
\item By training a model with sensory memory as input, an agent (e.g., agent 1) derives model parameters, encoding sensory memory into implicit (procedural) memory; \item To share knowledge with another agent and enable KD, agent 1 performs inference using its model parameters, yielding model outputs. This extracts concrete knowledge from implicit memory; 
\item To learn from agent 1, agent 2 uses local sensory memory and the concrete knowledge received from agent 1 to train its own model, thus updating its procedural memory. 
\end{itemize}

In another approach, an agent can directly share its model parameters to transfer knowledge to another agent, as shown in Fig. \ref{fig:kd2-b} for general models and Fig. \ref{fig:kd2-c} for knowledge graphs. 
\begin{itemize} 
\item After local training, an agent shares its model parameters, representing abstract knowledge, directly with agent 2. 
\item Agent 2 then stores the received model parameters as implicit memory and extracts concrete knowledge from this memory using inputs from its sensory memory. 
\item The concrete knowledge is subsequently used to guide agent 2's local training, achieving effective knowledge transfer. 
\end{itemize}
In the case of knowledge graphs, either inputs or model parameters can incorporate the graph structure. An agent can derive a knowledge graph from inputs, transforming sensory memory into semantic memory. This semantic memory can then be used as input to train agent models, as illustrated in Fig. \ref{fig:kd2-b}. Alternatively, the knowledge graph itself can function as a model, shared as abstract knowledge to support knowledge transfer, as shown in Fig. \ref{fig:kd2-c}.


\section{Rethinking KD for Different Multi-agent Collaborative Patterns}

In this section, we present the common collaborative patterns and their corresponding memory distributions among agents.  These patterns primarily include the distributed (centralized) pattern, the hierarchical pattern, and the decentralized pattern.

\begin{figure*}
	\centering
	\includegraphics[width=1\linewidth]{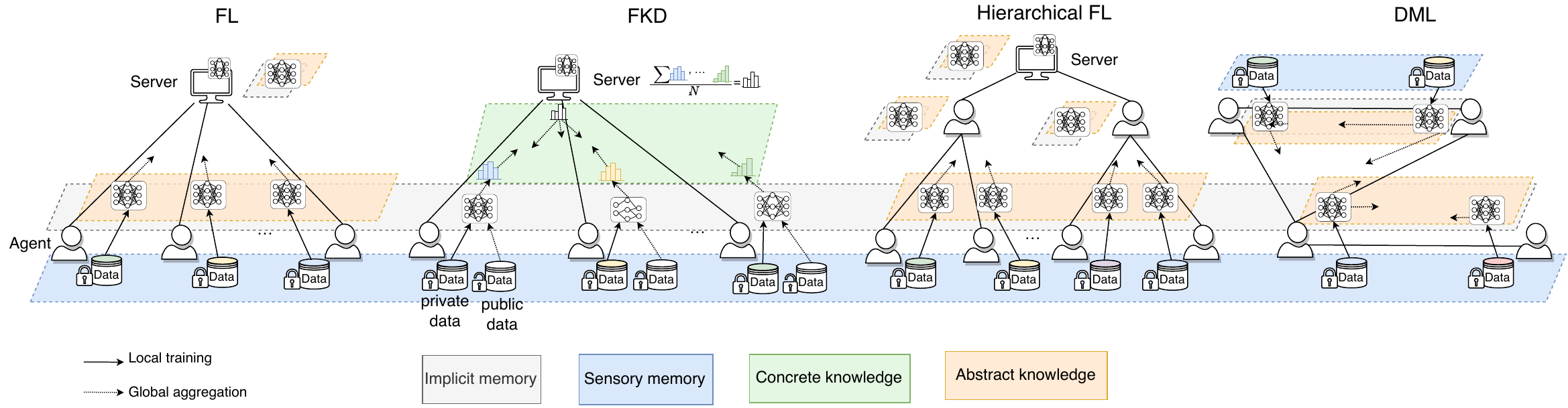}
	\caption{Collaborative patterns among agents.}
	\label{fig:pattern}
\end{figure*}

Different kinds of knowledge can be shared among agents to enhance collaborative learning, as shown in the FL and FKD in Fig. \ref{fig:pattern}, each offering distinct levels of privacy protection and communication efficiency. Concrete knowledge, such as model outputs, allows agents to retain a high degree of model privacy, as only the output predictions are shared rather than raw data or model parameters. This approach reduces communication overhead, as outputs are typically smaller in size than full model parameters. On the other hand, sharing abstract knowledge, such as model parameters or a knowledge graph, can enable deeper collaboration by directly transferring learned patterns and representations. However, this method may expose more details of an agent's model structure and training data characteristics, raising privacy concerns. Therefore, choosing the type of knowledge to share involves balancing privacy, communication costs, and the depth of knowledge transfer to achieve effective and secure collaborative learning. Table \ref{tab:com-KD} compares two types of KD methods used in collaborative learning—output sharing (black-box models) and parameter sharing.  Output sharing offers advantages in terms of lower communication and computation overhead, stronger privacy protection, and flexibility in handling heterogeneous models. However, it typically achieves lower accuracy compared to parameter sharing. On the other hand, parameter sharing can yield higher accuracy but comes with increased communication and computation costs, reduced privacy, and limited support for heterogeneous model architectures.
Next, we introduce various collaborative patterns, focusing on memory distribution and knowledge exchange within each pattern.

\begin{figure*}
	\centering
	\includegraphics[width=0.85\textwidth]{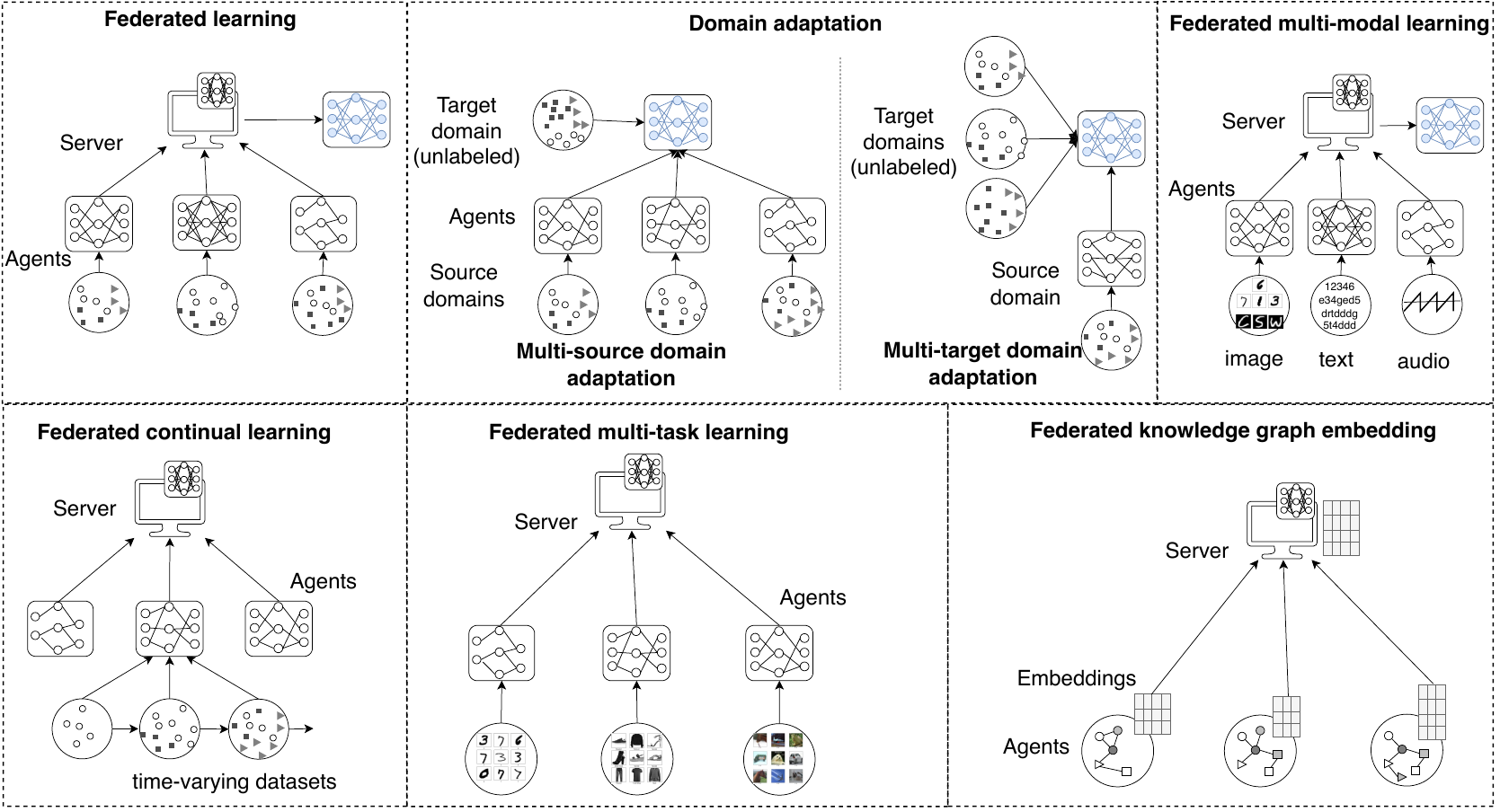}
	\caption{Frameworks of multi-agent collaborative learning under distributed pattern.}
	\label{fig:framework-distributed-pattern}
\end{figure*}

\begin{table*}[t] \centering
    \caption{Comparison of different KD methods for collaborative learning. }
    \label{tab:com-KD}
        \begin{center}
            \begin{small}
\begin{tabular}{p{5cm}p{5cm}p{3.5cm}}
\hline 
 & Output sharing (black-box models) & Parameter sharing \tabularnewline
\hline
Communication cost & low & high \tabularnewline
\rowcolor{gray!15} Computation cost & low & high \tabularnewline
Privacy & high  & low \tabularnewline
\rowcolor{gray!15} Accuracy & low & high \tabularnewline
Heterogeneous data & \cmark & \cmark \tabularnewline
\rowcolor{gray!15} Heterogeneous local models & \cmark & \xmark \tabularnewline
\hline
	\end{tabular}
\end{small}
\end{center}
\end{table*}

\subsection{Distributed (Centralized) Pattern}
The distributed pattern, as illustrated in Fig. \ref{fig:pattern}, is characterized by agents that are geographically dispersed and interconnected through communication networks. The agents in this context can be entities like corporations, organizations, or sensors that possess private datasets. In this pattern, a central server acts as an aggregator, coordinating the collaboration among these agents. Each agent retains its sensory memory locally, this includes private and public datasets, while sharing its implicit memory, such as model parameters and model outputs, with the central server. Different from the KD between two agents, the central server in distributed learning retains an additional implicit memory to facilitate model aggregation.

Following the distributed pattern, there are several problems of multi-agent collaborative learning according to the data and model characteristics as demonstrated in Fig. \ref{fig:framework-distributed-pattern}.
\begin{itemize}
    \item Federated learning (FL) \cite{mcmahan2017communication}: When data (with homogeneous or heterogenous distributions) are distributed across multiple agents (e.g., devices, institutions) and all agents intend to collaboratively train a model without sharing their local data, federated learning is a promising approach. Instead of sending data to a central server, each agent trains the model locally and only shares model updates (such as gradients or parameters), i.e., the encoded abstract knowledge from the data in sensory memory. The central server stores and updates the abstract knowledge for global model aggregation. This technique preserves data privacy and reduces the risk of data breaches, making it ideal for applications in healthcare, finance, and mobile devices. 
    
    To enhance privacy preservation, federated knowledge distillation (FKD) \cite{jeong2018communication} builds upon FL by sharing model outputs with the central server instead of model parameters. FKD relies on the use of a public dataset shared among agents for concrete knowledge extraction and supports heterogeneous model architectures, offering a higher degree of model privacy and greater flexibility in model deployment compared to FL. In FKD, the central server receives and aggregates the concrete knowledge of agents. In certain variations of FKD, such as \texttt{FedDF} \cite{lin2020ensemble}, the central server requires access to a public dataset as well as the agents' local models. Here, knowledge transfer among agents is conducted on the server, necessitating both sensory and implicit memory to facilitate this process.
    \item Multi-agent domain adaptation (MADA)\cite{farahani2021brief}: Domain adaptation is a subset of transfer learning where a model trained on one domain (source domain) is adapted to perform well on a different, but related domain (target domain). The goal is to transfer knowledge from the source domain, which has labeled data, to the target domain, where labeled data might be scarce or unavailable. 
    In multi-agent settings, domain adaptation can take two forms: multi-source domain adaptation, where knowledge from multiple source domains is transferred to a single target domain, and multi-target domain adaptation, where knowledge from a single source domain is transmitted to multiple target domains.
    \item Federated multi-modal learning (FML) \cite{lin2023federated}: Federated multi-modal learning extends the idea of FL to scenarios where agents have data from multiple modalities, such as text, images, audio, or sensor data. The goal is to collaboratively learn a model that can effectively process and integrate information from these different data types while respecting the privacy of each agent's data. This approach is beneficial in applications like healthcare, where patient data may include medical images, clinical text, and sensor readings.
    \item Federated continual learning (FCL) \cite{dong2022federated}: Federated continual learning is a method that combines federated learning with continual learning, where the model is continuously updated over time as new data arrives. In FCL, agents can continue to learn from their local data as it evolves, and these updates are shared with the global model. This allows the system to adapt to new tasks or changing data distributions without forgetting previously learned tasks (avoiding catastrophic forgetting). It is particularly useful in dynamic environments like IoT, where data streams over time.
    \item Federated multi-task learning (FMTL) \cite{NIPS2017_6211080f}: Federated multi-task learning allows different agents to collaboratively train models for different, but related, tasks. Instead of learning a single shared model, FMTL learns task-specific models while exploiting commonalities between tasks. Each agent can focus on its own specific task, but the knowledge from other agents helps improve the model for the individual task. This is useful when agents have varying objectives, such as personalized recommendation systems or domain-specific predictive models.
    \item Federated knowledge graph embedding (FKGE) \cite{chen2021fede}:  Knowledge graph embeddings (KGE) are vector representations of entities and relationships in a knowledge graph, which are used to capture the semantic and structural information of the graph in a lower-dimensional space. These embeddings can be used for various downstream tasks such as link prediction, node classification, and entity resolution. FKGE allows multiple agents to collaborate to learn the embeddings without sharing their actual knowledge graph data.
\end{itemize}
It is important to note that in MADA, FML, FCL, FMTL, and FKGE, we only introduce the case of abstract knowledge sharing. However, sharing concrete knowledge is also feasible through the use of KD, as illustrated by FKD and FL.
Depending on the type of knowledge being shared, different types of memory are maintained on the server, for instance, implicit memory to store model parameters and knowledge graphs and sensory memory to store public datasets.

\subsection{Hierarchical Pattern}
The hierarchical pattern \cite{lackinger2024inference}, as illustrated in Fig. \ref{fig:pattern}, offers an efficient and scalable topology for managing a large number of agents with distributed sensory memory by clustering them according to specific criteria. For instance, agents that are geographically proximate can be grouped into clusters, each managed by a local server. These local servers then communicate with a global server, which performs global model aggregation over the abstract knowledge stored as implicit memory. Additionally, in frameworks involving KD, agents that exhibit high similarity can be clustered to enhance knowledge transfer efficiency. In this hierarchical structure, agents retain their sensory memory, such as private data, locally while sharing their concrete knowledge. 

\subsection{Decentralized Pattern}
In the absence of a central server, agents can be organized in a decentralized manner to facilitate collaborative learning. In this decentralized pattern, agents communicate directly with each other in a peer-to-peer fashion. Compared to distributed and hierarchical patterns, the decentralized approach offers reduced communication overhead due to its sparse communication topology. Furthermore, this topology enhances security by eliminating the risk associated with a single point of failure, thereby increasing the system's robustness and resilience.
Similarly, in a decentralized collaborative pattern, agents can share either abstract or concrete knowledge, offering varying levels of privacy, resource costs, and model performance.


\section{KD empowered Memory Consolidation under Distributed Collaborative Pattern}
In this section, we review the use of KD for memory consolidation within distributed collaborative frameworks across various tasks. 

\subsection{Federated Learning}

The most widely adopted method for training a global machine learning model among distributed agents without sharing local data is FL \cite{mcmahan2017communication, kairouz2019advances}. Vanilla FL, such as federated averaging (\texttt{FedAvg}) \cite{mcmahan2017communication} trains models across agents over multiple rounds. Each round consists of the following steps:
\begin{itemize}
    \item Local training: Each client trains its local model using its own data and then uploads the trained model to the server.
    \item Global aggregation: The server aggregates the clients' local models into a global model according to predefined rules, such as weighted parameter averaging. The updated global model is then broadcast to the clients for the next round of training.
\end{itemize}
We review works on FL empowered by KD with a focus on model heterogeneity, data heterogeneity, resource heterogeneity, and privacy. Table \ref{tab:related-work-FKD} summarizes the literature on FL combined with KD.

\begin{table*}[t] \centering
    \caption{Literature of Federated Learning with Knowledge Distillation. }
    \label{tab:related-work-FKD}
        \begin{center}
 \begin{footnotesize}

  \begin{tabular}{p{2.2cm}p{0.9cm}p{0.9cm}p{0.9cm}p{0.8cm}p{0.8cm}p{0.8cm}p{0.8cm}p{0.8cm}p{1.0cm}p{3.2cm}}


		\hline
  Ref. & Hetero-geneous models  & Black-box  models & Hetero-geneous data & Has public data & Share average feature & Generate public data & Genera-lization  bound & Conver-gence& agent operation & Server operation \tabularnewline
  \hline
  \texttt{FedMD} \cite{li2019fedmd}  & \cmark & \cmark & \xmark & \cmark  & \xmark & \xmark& \xmark & \xmark & KD & output averaging\tabularnewline
  \rowcolor{gray!15}\texttt{DS-FL} \cite{itahara2021distillation} & \cmark & \cmark & \cmark & \cmark & \xmark& \xmark & \xmark & \xmark & KD & output averaging\tabularnewline
  \texttt{Cronus} \cite{hongyan2019cronus} & \cmark & \cmark & \xmark & \cmark & \xmark & \xmark & \xmark & \xmark & KD &output averaging\tabularnewline
  \rowcolor{gray!15}\texttt{FCCL+} \cite{huang2023generalizable} & \cmark & \cmark &\cmark &\cmark  & \xmark & \xmark & \cmark & \xmark  & KD & output averaging, intermediate feature similarity averaging\tabularnewline
  \texttt{FedCKT} \cite{cho2021personalized}  & \cmark & \cmark & \cmark & \cmark & \xmark & \xmark& \cmark & \cmark & KD & clustering \tabularnewline
  \rowcolor{gray!15}\texttt{CFD} \cite{sattler2020communication}  & \cmark & \cmark & \xmark& \cmark & \xmark  &\xmark & \xmark & \xmark & KD & model training \tabularnewline
  \texttt{MHAT} \cite{hu2021mhat}   & \cmark & \cmark & \xmark & \cmark & \xmark& \xmark & \xmark & \xmark & KD & model training  \tabularnewline
  \rowcolor{gray!15}\texttt{FedGems} \cite{cheng2021fedgems} & \cmark & \cmark & \xmark & \cmark & \xmark & \xmark & \xmark & \xmark & KD & model training\tabularnewline
  \texttt{FedAL} \cite{10606337} & \cmark & \cmark & \cmark  & \cmark & \xmark & \xmark & \cmark & \cmark & KD  & output averaging, discriminator\tabularnewline 
  \rowcolor{gray!15}\texttt{FedDF} \cite{lin2020ensemble} & \cmark & \xmark &\xmark & \cmark & \xmark & \xmark & \cmark & \xmark & - & parameter averaging, KD \tabularnewline
  \texttt{FedAUX} \cite{sattler2021fedaux}  & \cmark & \xmark & \xmark & \cmark & \xmark & \xmark & \xmark & \xmark & KD & parameter averaging \tabularnewline
  \rowcolor{gray!15}\texttt{FedBKD} \cite{lee2020edge}& \cmark & \xmark & \cmark & \cmark & \xmark  & \xmark  & \xmark  & \xmark  &KD & KD \tabularnewline
 \texttt{FD}/
 \texttt{HFD} \cite{jeong2018communication, 8904164}  & \cmark & \cmark & \cmark & \xmark & \cmark &\xmark & \xmark & \xmark & KD & output averaging\tabularnewline
 \rowcolor{gray!15}\texttt{FedProto} \cite{tan2021fedproto}  & \cmark & \cmark & \cmark & \xmark & \cmark  &\xmark & \xmark & \xmark & KD & output averaging\tabularnewline
\texttt{FedHKD} \cite{chen2023best} & \cmark & \cmark & \xmark & \xmark & \cmark & \xmark & \xmark & \cmark  & KD & output averaging\tabularnewline
\rowcolor{gray!15}\texttt{FedGKT} \cite{he2020group}  & \cmark  & \cmark & \xmark & \xmark & \cmark & \xmark  &  \xmark & \xmark & KD & global classifier\tabularnewline
\texttt{FedDKC} \cite{wu2024exploring} & \cmark & \cmark & \xmark & \xmark & \cmark & \xmark & \xmark & \xmark & KD & global classifier, KD\tabularnewline
\rowcolor{gray!15}\texttt{ADCOL} \cite{pmlr-v202-li23j} & \cmark & \cmark & \cmark  & \xmark & \cmark & \xmark & \cmark & \xmark & KD  & discriminator\tabularnewline
\texttt{FedZKT} \cite{zhang2021fedzkt}  & \cmark  & \xmark  & \xmark & \xmark & \xmark & \cmark  & \xmark  & \xmark & - & generator, KD\tabularnewline
\rowcolor{gray!15}\texttt{FedFTG} \cite{9879661} & \cmark & \xmark &\cmark& \xmark & \xmark & \cmark & \xmark & \xmark & - & generator, KD \tabularnewline
\texttt{FedGD} \cite{zhang2023towards} & \cmark & \cmark & \cmark  & \xmark & \xmark & \cmark & \xmark & \xmark & KD, discriminator & output averaging, generator\tabularnewline 
\rowcolor{gray!15}\texttt{FedGen} \cite{zhu2021data} & \xmark & \xmark& \cmark & \xmark & \xmark & \cmark & \cmark &  \xmark & KD & parameter averaging, generator \tabularnewline
\texttt{FedBE} \cite{chen2020fedbe} & \xmark & \xmark & \cmark & \cmark & \xmark & \xmark & \xmark & \xmark & KD &  ensemble model construction, parameter averaging\tabularnewline
\rowcolor{gray!15}\texttt{FedCAD} \cite{he2022class} & \xmark & \xmark & \cmark 
 & \cmark  & \xmark & \xmark & \xmark & \xmark & KD & parameter averaging\tabularnewline
\texttt{FedSSD} \cite{he2022learning} & \xmark & \xmark  & \cmark  & \cmark & \xmark  & \xmark  & \xmark & \cmark & KD & parameter averaging  \tabularnewline
\rowcolor{gray!15}\texttt{FedMLB} \cite{pmlr-v162-kim22a} & \xmark & \xmark  & \cmark  & \xmark   & \xmark   & \xmark   & \xmark  & \xmark & KD & parameter averaging \tabularnewline
\texttt{DaFKD} \cite{wang2023dafkd}& \xmark & \xmark & \cmark  & \xmark & \xmark & \cmark & \cmark & \xmark & KD, generator, discriminator & weighted output averaging\tabularnewline 
\rowcolor{gray!15}\texttt{QuPed} \cite{ozkara2021quped} & \xmark & \xmark & \cmark& \xmark& \xmark& \xmark& \xmark   & \cmark  & KD & parameter averaging \tabularnewline
\hline
	\end{tabular}

\end{footnotesize}
\end{center}
\end{table*}

\subsubsection{Model Heterogeneity}
Most FL algorithms, e.g., \texttt{FedAvg}, rely on distributed stochastic gradient descent (SGD) and assume that all agents use identical model architectures. However, in real-world scenarios, agents, such as organizations, may design their own unique model architectures, making parameter averaging, as done in \texttt{FedAvg}, impractical. 
One advantage of incorporating KD in federated settings is its flexibility in handling heterogeneous models across agents. KD operates on model outputs, which are standardized to the same dimensions across clients, regardless of differing model architectures.

Typical KD-enhanced FL algorithms follow these steps:
\begin{itemize}
    \item Local training: Each client trains its local model on private data and uploads either the local model or its outputs on a common dataset to the server. 
    \item Global aggregation: The server collects the local model outputs (either by running the models on a common dataset or receiving them directly from the clients) and performs global aggregation, e.g., output averaging, to aggregate the clients' knowledge.
    \item Local KD: Using the aggregated global knowledge, clients'  local models are re-trained through KD to align them with the global knowledge. This re-training can occur on the server or locally. If conducted on the server, the server re-trains the models and sends them back to the clients. Alternatively, the server can broadcast the globally aggregated knowledge to facilitate client-side KD.
\end{itemize}
Different methods of local knowledge extraction can be used to support KD, including the use of an unlabeled public dataset, private datasets, or a generated public dataset. We summarize the relevant literature below.

\textbf{Extracting knowledge using unlabeled public dataset.}
Collecting an unlabeled public dataset is significantly easier than obtaining a labeled one, as it avoids the labor-intensive task of data labeling. Agents can contribute portions of their local data to help construct the public dataset. Alternatively, data with similar characteristics to the clients' local datasets can be used for knowledge transfer. For instance, CIFAR-10 can serve as a public dataset for CIFAR-100~\cite{lin2020ensemble}.

\begin{figure}
	\centering
	\includegraphics[width=0.9\columnwidth]{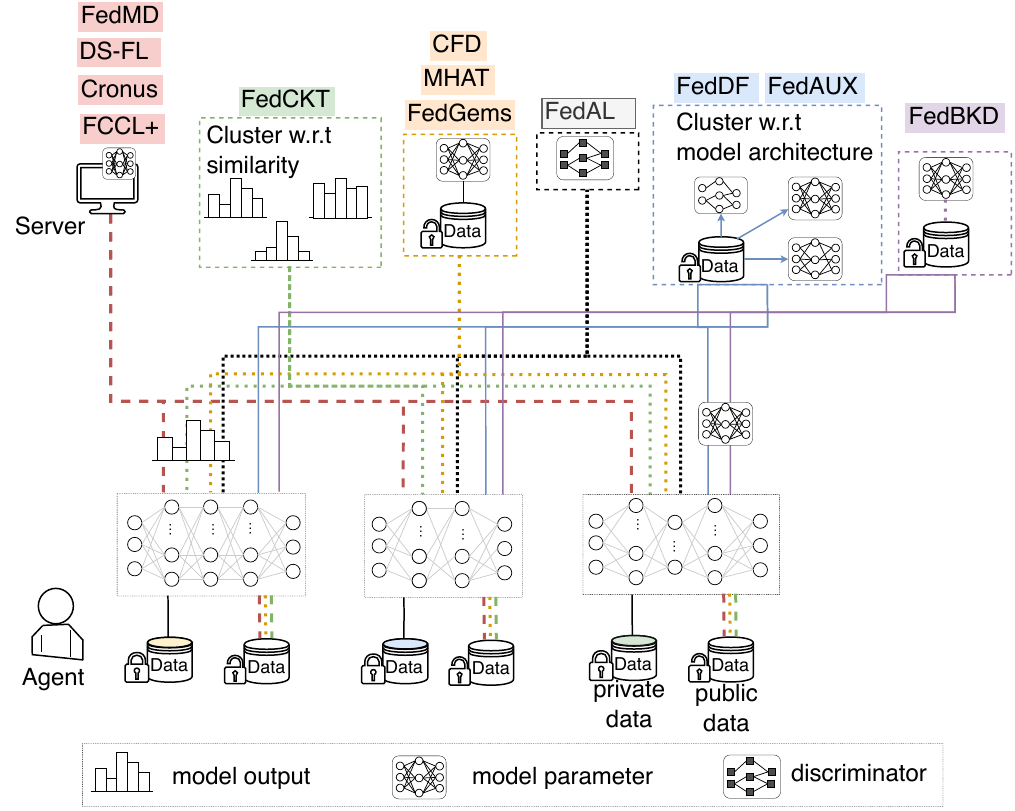}
	\caption{Framework of methods tackling heterogeneous model architectures using unlabeled public dataset.}
	\label{fig:model-heter-public}
\end{figure}

Figure \ref{fig:model-heter-public} illustrates methods for addressing model heterogeneity using an unlabeled public dataset. To leverage this public dataset for knowledge transfer among agents, each agent can share its local model outputs on the public data with the server without revealing its local model parameters. The server can then perform global aggregation through various methods, including output averaging, clustering, or global model training.
In \texttt{FedMD} \cite{li2019fedmd}, \texttt{DS-FL} \cite{itahara2021distillation}, \texttt{Cronus} \cite{hongyan2019cronus}, and \texttt{FCCL+}  \cite{huang2023generalizable}  the server averages the model outputs and sends the aggregated output back to all agents. Specifically, \texttt{FCCL+} replaces the agent-side KD in \texttt{FedMD} with a correlation-based method that measures the relationship between the local model outputs and the global averaged outputs on the public dataset. This method introduces feature-level knowledge exchange, enabling agents to learn more diverse information from each other, thus improving generalization.
In \texttt{FedCKT}~\cite{cho2021personalized}, the server clusters agents based on the similarity of their local model outputs, allowing agents to select similar peers for KD, which supports personalized model training for each agent.
In \texttt{CFD}\cite{sattler2020communication} and \texttt{MHAT}\cite{hu2021mhat}, the server trains a global model using the predictions provided by agents on the public dataset and regenerates global predictions to facilitate KD on the client side.
In \texttt{FedGems}~\cite{cheng2021fedgems}, the server trains a global model using agents' local model outputs and selects specific public data to efficiently transfer knowledge from the global model to the agents' local models. \texttt{FedAL} \cite{10606337} trains a discriminator on the server to differentiate agents' local model outputs over the public dataset, encouraging agents to produce consistent output probabilities.

To reduce the computational overhead for agents, the server can perform both global aggregation and KD. In this case, clients share their local model parameters with the server. For instance, both \texttt{FedDF} \cite{lin2020ensemble} and \texttt{FedAUX}\cite{sattler2021fedaux} combine the approaches of \texttt{FedAvg} and \texttt{FedMD} by incorporating parameter averaging and output averaging. On the server side, agents' local models are clustered based on their architecture, and parameter averaging is applied within clusters of models with the same architecture. For different clusters, KD is used to transfer knowledge across them.
Additionally, \texttt{FedBKD}~\cite{lee2020edge} trains a global model on the server and uses KD to facilitate bidirectional knowledge transfer between the global model and the agents' local models.


\textbf{Extracting knowledge using private dataset.}
When public data is not available among agents, each agent can leverage its local data to extract knowledge from its local model. However, directly sharing local data features poses a risk of privacy leakage. An alternative approach is to share the average local model output/feature over the local data, which effectively reflects the distribution of that data while maintaining privacy.

Figure \ref{fig:model-heter-private} illustrates methods for addressing model heterogeneity using clients' private datasets, where various forms of average features are shared to facilitate global aggregation and knowledge transfer.
In \texttt{FD}\cite{jeong2018communication,park2019distilling} and \texttt{HFD}\cite{8904164}, each agent computes the average local model output for each class of local data and shares these averages with the server for KD.
In \texttt{FedProto}\cite{tan2021fedproto} and \texttt{FedHKD}\cite{chen2023best}, the local model is divided into two segments: the feature extractor and the classifier. Agents share their class-wise average intermediate features, obtained after the extractor, with the server, which helps update the feature extractors to align more closely with the global class-wise average intermediate features.
In \texttt{FedGKT}~\cite{he2020group}, agents share their intermediate features along with the corresponding ground-truth labels with the server. The server aggregates the agents' features by training a global classifier, thereby achieving global knowledge aggregation. The global model outputs from the server are then fed back to the agents for KD.
In \texttt{FedDKC}~\cite{wu2024exploring}, both the intermediate features and model outputs of agents' local models are uploaded to the server for global knowledge aggregation. Additionally, the ground-truth labels of the input data are shared with the server to aid in training a global classifier.
In \texttt{ADCOL}~\cite{pmlr-v202-li23j}, the server trains a discriminator to differentiate between agents' local model outputs, facilitating the production of consistent intermediate features across all agents.


\begin{figure}
	\centering
	\includegraphics[width=0.9\columnwidth]{figure/fig_fd.pdf}
	\caption{Overall design of existing approaches tackling model architecture heterogeneity using private datasets.}
	\label{fig:model-heter-private}
\end{figure}

\textbf{Extracting knowledge using generated public dataset.}
When agents are unwilling to share their local features extracted from local data, and public data is unavailable, generating public data becomes desirable.
In \texttt{FedZKT}\cite{zhang2021fedzkt} and \texttt{FedFTG}\cite{9879661}, the server collects all agents' local models to perform model aggregation. Specifically, the server trains a generator based on the collection of local models and uses this generator to train a global model. KD is then applied to transfer the knowledge from the global model to each local model.
\texttt{FedGD}~\cite{zhang2023towards} also trains a unified generator on the server to facilitate KD among agents. Each agent trains a discriminator to enhance the quality of the generated data.

In one-shot FL, \texttt{FedCVAE}~\cite{heinbaughdata} utilizes local models to generate global data, which is subsequently used to train a global model.
Finally, \texttt{DENSE}~\cite{zhang2022dense} addresses model architecture heterogeneity in one-shot FL by training a generator to create data for model distillation from the ensemble of local models to the final global model.

\subsubsection{Data Heterogeneity.}
Data heterogeneity among agents poses a significant challenge in collaborative learning, as it hinders the efficiency of knowledge transfer among agents. Existing research has primarily addressed data heterogeneity from the following perspectives: adversarial learning, data augmentation, using history models, new aggregation methods, less forgetting, and personalization.

\textbf{Data Augmentation.}
To address the challenge of non-IID data distribution among agents, a straightforward approach is to generate new data to reduce data heterogeneity.
In \texttt{FD}~\cite{jeong2018communication}, agents identify the missing labels in their local datasets and upload a few data samples to the server. The server then trains a generator using a conditional GAN, which is broadcast to all agents. The agents use the generator to supplement their local datasets with the missing target labels and train their models on this augmented data to enhance generalization. The augmented data helps reduce the bias in the agents' local data, leading to improved global model performance.
Similarly, in \texttt{FedGen}~\cite{zhu2021data}, the server learns a generator that captures the global data characteristics and shares it with all agents.
\texttt{FedFTG} \cite{9879661} and \texttt{FedGD} \cite{zhang2023towards} generate data samples that reflect global data characteristics, enabling effective KD among agents with heterogeneous data distributions.

\textbf{Private knowledge sharing.}
Agents can share the knowledge extracted from their private local data to help compensate for data heterogeneity among them.
In FedProto~\cite{tan2021fedproto}, agents share their class-wise average intermediate features with the server. The server then aggregates these features, and agents update their feature extractors to align more closely with the global class-wise average intermediate features, thereby mitigating data heterogeneity across agents.

\textbf{New aggregation methods.}
Some works propose novel aggregation methods that account for the drift among agents due to data heterogeneity.
For example, \texttt{DS-FL} \cite{itahara2021distillation} uses output averaging on the server to aggregate agents' local model outputs. 
\texttt{FedBKD}~\cite{lee2020edge} utilizes KD to aggregate agents' local knowledge into a global model instead of relying on parameter or output averaging, allowing the server to effectively mimic all agents.
\texttt{FedBE}~\cite{chen2020fedbe} fits a Gaussian or Dirichlet distribution to local models, constructs a Bayesian model ensemble, and uses KD to transfer this ensemble knowledge to the global model, enhancing performance under non-IID data.
In \texttt{FedCAD}\cite{he2022class}, a public dataset assesses the inference confidence for each data class, which informs class-wise weighting for local training and KD on the agent side. Similarly, \texttt{FedSSD}\cite{he2022learning} leverages a labeled public dataset to measure the credibility of the global model for each data class, enabling agents to update local models and prevent forgetting of global knowledge.
Finally, \texttt{FedMLB}~\cite{pmlr-v162-kim22a} introduces multiple auxiliary model branches for each local model, allowing them to learn from the global model and improve robustness simultaneously.

\textbf{Adversarial learning.} 
Adversarial learning (AL) is a machine learning approach that focuses on training models to be robust against adversarial examples—inputs that are intentionally designed to deceive the model into making incorrect predictions. This typically involves two main components: a generator that creates adversarial examples and a discriminator (or classifier) that learns to distinguish between regular and adversarial inputs. The generator aims to improve its ability to create effective adversarial samples, while the discriminator aims to correctly classify these samples. This interplay helps to more resilient models toward non-IID data among agents. 

Some works on standard (non-federated) KD have demonstrated the effectiveness of AL in improving the accuracy of student models, particularly in the scenario of one teacher and one student model sharing the same training dataset~\cite{xu2017training, liu2020learning, gao2020private, wang2018kdgan, wang2018adversarial}. However, these approaches rely on the original labeled training data used by the teacher model.
In the federated setting, \texttt{ADCOL}~\cite{pmlr-v202-li23j} addresses the challenge of non-IID data across agents by employing adversarial learning between agents. A discriminator is trained on the server to help all agents produce a consistent representation distribution.
Additionally, \texttt{DaFKD}~\cite{wang2023dafkd} trains a generator on each agent and aggregates the local generators on the server to support KD. Each agent also trains a domain discriminator to differentiate between generated and local data. Then, a correlation factor is obtained to weigh agents for global output averaging on the server.
\texttt{FedAL} \cite{10606337}  trains a discriminator on the server to differentiate agents' local model outputs, encouraging agents to produce consistent output probabilities.

\textbf{Less forgetting.}
Less forgetting refers to techniques designed to prevent a model from ``forgetting" previously learned knowledge when it is trained on new data or tasks. The phenomenon of a model losing its ability to perform previous tasks after learning new ones is known as catastrophic forgetting. This issue is a central challenge in lifelong learning and continual learning (or incremental learning), where a model is exposed to a sequence of tasks and must learn new information without erasing its performance on prior tasks. 
\begin{itemize}
    \item Lifelong learning (LLL) \cite{li2017learning,aljundi2018memory,yao2019adversarial}: LLL try to train a model that performs well on a sequence of tasks incoming separately and sequentially with only the most recent data. To alleviate the forgetting in LLL, \texttt{LwF} \cite{li2017learning} constrains model updates by limiting the divergence between the new dataset and the previous one through KD. \texttt{MAS} \cite{aljundi2018memory} evaluates the importance of model parameters and retains the most critical ones during training on new data. \texttt{AFA} \cite{yao2019adversarial} employs adversarial learning to identify shared model parameters between the old and new tasks.
    \item Continual learning (CL): CL, also known as incremental learning, focuses on progressively learning from new tasks while retaining knowledge from previous ones. Many related works, such as \texttt{LwM} \cite{dhar2019learning}, \texttt{iCaRL} \cite{8100070}, \texttt{EEIL} \cite{castro2018end}, \cite{jung2018less}, and \texttt{Online EWC} \cite{schwarz2018progress}, leverage the trained model parameters from previous tasks to constrain the updates for new tasks, ensuring compatibility with both old and new tasks. \texttt{EWC} \cite{kirkpatrick2017overcoming} and its extension \texttt{EWC}++ \cite{chaudhry2018riemannian} compute the Fisher information matrix to assess the importance of model parameters, enabling weighted updates. \texttt{GeoDL} \cite{Christian2021MGeoCont} and \cite{hou2019learning} use cosine distance to limit updates based on intermediate model features and outputs instead of solely on model parameters. For a comprehensive survey on incremental learning, see \cite{liu2023incremental}.
\end{itemize}

In a federated setting, agents may forget the knowledge from their local datasets / global knowledge during global aggregation / local training, making less-forgetting approaches crucial. Various models serve as benchmarks to distill historical knowledge and mitigate forgetting.
In \texttt{FedGKD} \cite{yao2021local}, the server maintains multiple \textit{history local models} to extract improved knowledge, thereby alleviating drift among agents' local datasets and enhancing overall model accuracy. \texttt{pFedSD} \cite{9964434} emphasizes the retention of each agent's self-knowledge during model aggregation, employing KD between the current local model and the one from the previous round to train personalized models for each agent. Additionally, \texttt{FCCL} \cite{huang2022learn} first trains \textit{optimal local models} for agents and maintains a cross-correlation matrix to learn a generalizable representation amid domain shifts. Both inter- and intra-domain KD are proposed to further prevent forgetting. Furthermore, \texttt{FedCurv} \cite{shoham1910overcoming} enables agents to share the \textit{Fisher information matrix} of their local models to regularize their local model updates.

To prevent the forgetting of global knowledge during local training, the \textit{global model obtained in the previous training round} is used to extract this knowledge. For instance, \texttt{FedSSD} \cite{he2022learning} employs a labeled public dataset to assess the credibility of the global model for each class of data. This credibility helps agents update their local models, mitigating the loss of global knowledge.
\texttt{FedLSD} \cite{lee2021preservation} reduces forgetting of the global model by distilling knowledge extracted from the global model using agents' private datasets, ensuring that agents' local models remain aligned with the global model while maintaining privacy.
\texttt{FedReg} \cite{xu2022acceleration} creates a perturbed dataset for each agent's local dataset to enhance the retention of knowledge from other agents and improve the robustness of model training. This perturbed dataset consists of data points with significant changes in predicted values.
\texttt{FCCL+} \cite{huang2023generalizable} addresses catastrophic forgetting during the local updating stage through non-target distillation, preserving interdomain knowledge while avoiding optimization conflicts.

To prevent forgetting of both global knowledge and knowledge from other agents, the \textit{global model and local models} can be utilized. For example, the authors in \cite{kim2024federated} focus on scenarios where each agent in FL has only a subset of classes. The non-IID data among agents can lead to catastrophic forgetting in FL. To tackle this challenge, KD is applied to both the global and local models to enhance robustness. In this approach, each agent shares part of its local data with the global model, and all agents exchange their local models with one another.
\texttt{FedBR} \cite{liu2023adaptive} introduces block-wise regulation and KD techniques to help agents absorb more global knowledge during local training. The server can either send the entire global model or specific blocks from the latest global model to an agent in each training round, improving the training performance of the global model in a resource-efficient manner. This block-wise regularization employs self-knowledge distillation, where each agent uses the final model output to guide updates to its shallow layers.
\texttt{Flashback} \cite{aljahdali2024flashback} emphasizes forgetting during both local training and global aggregation, proposing a KD method with dynamic weights across classes for both stages. \texttt{DFRD} \cite{wang2024dfrd} enhances training of the generator for producing synthetic data for KD on agents. \texttt{FedRAD} \cite{tang2023fedrad} applies KD to train both local and global models on each agent, alleviating model forgetting through rational KD over batches of data, facilitating knowledge transfer. \texttt{FedNTD} \cite{lee2106preservation} addresses local model forgetting of global knowledge by minimizing the KL divergence between global and local model outputs on non-true classes, achieved by exchanging model parameters between agents and the server.

\textbf{Personalization.}
Building on the heterogeneous data distributions of agents, personalization refers to techniques that tailor the global model to meet the specific needs of each agent. Traditional FL approaches aim to train a single global model by aggregating model updates from all clients, which can perform suboptimally for individual clients if their data is significantly different (i.e., non-IID). Personalization methods address this by allowing each client to benefit from FL while still having a model that fits their unique data characteristics.
For example, \texttt{FedCKT}~\cite{cho2021personalized} clusters agents based on the similarity of their model output probability distributions, which reflect local data characteristics. Co-distillation is then applied within clusters to enhance personalization.
\texttt{FedHKD}~\cite{chen2023best} trains personalized models for agents using output averaging to facilitate knowledge transfer.
In \texttt{QuPed}~\cite{ozkara2021quped}, the server aggregates agents' local models via parameter averaging and broadcasts the global model back to agents. Each agent then extracts global knowledge from this model to update its local model, achieving a personalized adaptation.

\begin{table*}[h!] \centering
    \caption{Related works on less-forgetting of FL and FCL with KD. }
    \label{tab:related-work-less-forgetting}
        \begin{center}
            \begin{footnotesize}
 \begin{tabular}{p{2.3cm}p{2.8cm}p{3cm}p{3cm}p{3cm}}
\hline 
Ref. & Scenario & Loss function & Operating dataset & Refereed model\tabularnewline
\hline 
\texttt{FedGKD}~\cite{yao2021local} & FL & KD between current and history local models & local data & History local models\tabularnewline 
\rowcolor{gray!15} \texttt{pFedSD}  \cite{9964434} & FL & KD between current and previous local model outputs & local  data & previous local model \tabularnewline
 \texttt{FCCL} \cite{huang2022learn} & FL & KD between local and global model outputs and between current and the optimal local model outputs & local data & optimal local model, previous global model \tabularnewline
\rowcolor{gray!15} 
\texttt{FedCurv} \cite{shoham1910overcoming}  & FL & FIM weighted parameter change & local data & (FIM of) other agents' local models\tabularnewline
\texttt{FedSSD} \cite{he2022learning} & FL &  distance of class-weighted local and global model outputs & labeled public data, local data & previous global model\tabularnewline
\rowcolor{gray!15} 
\texttt{FedLSD} \cite{lee2021preservation} & FL & KD between local and global model outputs & local data & previous global model\tabularnewline
\texttt{FedReg} \cite{xu2022acceleration}  & FL & local training loss & sampled local data 
& previous global model \tabularnewline
\rowcolor{gray!15} 
\texttt{FCCL+}  \cite{huang2023generalizable} & FL & KD between local the global model outputs & local data & previous global model\tabularnewline 
\cite{kim2024federated} & FL & KD between local and global model outputs and others' local model output & unlabeled  local data & local and global models  \tabularnewline
\rowcolor{gray!15}  
\texttt{FedBR} \cite{liu2023adaptive} & FL & self-KD & local data & local model augmented by the global model \tabularnewline 
\texttt{Flashback} \cite{aljahdali2024flashback} & FL & KD between local and global model outputs & local and public data & local and global model\tabularnewline
\rowcolor{gray!15} 
 \texttt{DFRD}  \cite{wang2024dfrd} & FL & KD between local and global model outputs& local data and generator & local and global model\tabularnewline
 \texttt{FedRAD} \cite{tang2023fedrad} & FL & KD between local and global model outputs & local data & local and global model\tabularnewline
\rowcolor{gray!15}  
\texttt{FedNTD} \cite{lee2106preservation} & FL & KD between local and global model outputs & local data & local and global model \tabularnewline
 \cite{Fedrecon2021}& class-incremental FL & KD between local and global model outputs & local data & previous global model \tabularnewline
\rowcolor{gray!15} 
\cite{dong2022federated} & class-incremental FL & KD between current and previous model outputs & local data & previous global model  \tabularnewline
\texttt{FL-IIDS} \cite{jin2024fl}& class-incremental FL & KL between local and  global model outputs & replay buffer augmented local dataset & previous global model  \tabularnewline
\rowcolor{gray!15}  
\texttt{FedET} \cite{liu2023fedet} & class-incremental FL & KD between local and the selective history model outputs& replay buffer & aggregated history models\tabularnewline
\texttt{LGA} \cite{dong2023no} &  class-incremental FL &  KD between current and previous global model outputs & replay buffer &  previous global model\tabularnewline
\rowcolor{gray!15} 
\texttt{TARGET} \cite{zhang2023target} & class-incremental FL & KD between local and global model outputs & generator 
& previous global model \tabularnewline
\texttt{MFCL} \cite{babakniya2024data} &  class-incremental FL & local training loss & new and generated data & previous global model \tabularnewline
\rowcolor{gray!15} 
\texttt{FCILPT} \cite{liu2023federated} & class-incremental FL & cross-entropy on the most relevant prompt & local data and global prompts & -  \tabularnewline
\texttt{Re-Fed} \cite{li2024towards} & class and domain-incremental FL & local training loss & mix of new data and replayed data & - \tabularnewline
\rowcolor{gray!15} 
\texttt{FedPMR} \cite{wang2023federated} & task-incremental FL & MSE of local and global model outputs & replay buffer  & previous global model \tabularnewline
 \texttt{CFeD} \cite{ma2022continual} & task-incremental FL& KL of local and global model outputs&  unlabeled public dataset  & previous global model, local models \tabularnewline
\rowcolor{gray!15} 
\texttt{TRINA} \cite{li2024facing} & task-incremental FL & distance between the model outputs of local and generated data & generator  & - \tabularnewline
 \texttt{DAG} \cite{carta2023projected}& task-incremental FL & KD between local and global model outputs & out-of-distribution dataset & previous global model \tabularnewline
\hline 
	\end{tabular}
\end{footnotesize}
\end{center}
\end{table*}

\subsubsection{Resource Heterogeneity}
Resource heterogeneity refers to the variation in computational power, memory, storage capacity, and network bandwidth among different agents participating in the training process. This heterogeneity can significantly impact the training dynamics and overall performance of the federated learning system. 
In this subsection, we review the works of KD-empowered FL with computation and communication resource heterogeneity, as shown in table \ref{tab:related-work-resource-heter}.  

\textbf{Computation heterogeneity.}
Agents in federated learning may possess varying hardware capabilities, such as differing CPU and GPU specifications, leading to significant disparities in their computational performance. To address these heterogeneous computational resources, \texttt{DepthFL} \cite{kim2023depthfl} proposes pruning local models to better match each agent's local computation capacity.
\texttt{ScaleFL} \cite{ilhan2023scalefl} determines the reduction ratio for each agent based on their local resource budget and the global model's cost, allocating model components according to this ratio. Self-distillation is then applied to enhance the local model performance for agents with these heterogeneous model splits.
\texttt{FSMKD} \cite{luo2024federated} emphasizes the constraints posed by limited computational resources and employs split learning to mitigate computation overhead. In this approach, each client has a model head, tail, and a small body, while the server trains a larger model body. KD is performed on the outputs from both the clients' and the server's bodies to facilitate effective learning.

\textbf{Communication heterogeneity.}
In collaborative learning among distributed agents, the speed and reliability of internet connections can vary significantly. Slow or unstable connections can delay the transmission of model updates, resulting in inefficiencies during training.
To reduce the communication overhead for agents, \cite{gong2022preserving} quantizes the locally computed logits, enabling a communication-efficient FKD.
The authors in \cite{wu2022communication} propose a communication-efficient FL method through KD, where communication costs are minimized by exchanging local updates of a smaller model instead of the original large models.
In \texttt{CFD} \cite{sattler2020communication}, the authors address the heterogeneous communication capacities of agents and introduce soft-label quantization and delta-coding techniques to enhance the communication efficiency of FKD.

Considering both computation and communication resource heterogeneities, \texttt{FDL} \cite{zhang2023distill} integrates FL and FKD. Agents have the option to upload either local model parameters (as in FL) or model outputs (as in FKD), with the server aggregating the global model from both FL and FKD contributions. Additionally, they propose a dynamic selection algorithm for FL and FKD to minimize overall training time while accounting for both computation and communication constraints.
\texttt{FedBR} \cite{liu2023adaptive} introduces block-wise regulation and KD techniques that enable agents to absorb more global knowledge during local training. The server can decide to send the entire global model or specific blocks from the most recent global model to an agent during each training round, enhancing the training performance of the global model in a resource-efficient manner. When an agent receives either the complete or partial global model, it updates its local model by replacing parts of it with the global model and applying self-KD to further improve its performance. This approach effectively minimizes overall training time while adhering to loss values and the computation and communication resource capacities.

\begin{table*}[t] \centering
    \caption{Related works on KD empowered Federated Learning with resource heterogeneity. }
    \label{tab:related-work-resource-heter}
        \begin{center}
            \begin{small}
 \begin{tabular}{p{2.3cm}p{2.8cm}p{3.5cm}p{3cm}p{3cm}}
\hline 
Ref. & Considered resources & Agent & Server & Shared knowledge \tabularnewline
\hline
\cite{kim2023depthfl} & computation & pruned local model training, KD &  global model training & pruned local model / global model \tabularnewline
\rowcolor{gray!15} 
\texttt{ScaleFL} \cite{ilhan2023scalefl} & computation & split local model training & global model aggregation & local / global models\tabularnewline
\texttt{FSMKD} \cite{luo2024federated} & computation & small local model training & large model body training & model intermediate features / gradients \tabularnewline
\rowcolor{gray!15}
\cite{gong2022preserving} & communication & local model training, KD & output averaging & quantified the local logits / -(one-shot FL) \tabularnewline
\cite{wu2022communication} & communication & local model training, model decomposition, KD & global aggregation & small local/global model \tabularnewline
\rowcolor{gray!15}
\texttt{CFD} \cite{sattler2020communication} & communication & local model training, KD & output averaging & quantified the local logits \tabularnewline
\texttt{FDL} \cite{zhang2023distill} & computation \& communication & local model training, KD & parameter/output averaging & model parameters / outputs\tabularnewline
\rowcolor{gray!15}
\texttt{FedBR} \cite{liu2023adaptive}& computation \& communication & local model training, KD & parameter averaging & model segments\tabularnewline
\hline 
	\end{tabular}
\end{small}
\end{center}
\end{table*}

\subsubsection{Privacy}
Privacy preservation in collaborative learning is crucial as it ensures that sensitive data remains secure while enabling the sharing of insights and model improvements across diverse agents. In collaborative settings, participants often possess valuable and confidential information, such as personal data or proprietary business metrics. By implementing privacy-preserving techniques, such as differential privacy, secure multiparty computation, and federated learning, organizations can facilitate cooperation among agents without exposing individual data. This not only builds trust among participants but also complies with legal regulations, such as GDPR or HIPAA, which mandate stringent data protection standards. Ultimately, privacy preservation is essential for fostering a collaborative environment that encourages innovation while safeguarding personal and sensitive information.
Several works utilize KD for enhancing privacy preservation in FL \cite{li2020practical}. A comprehensive survey on privacy-preserving techniques in FL can be found in \cite{shao2023survey}, which reviews the literature related to the sharing of data, knowledge, and models between agents and the server in FL. In this subsection, we review the works on privacy reservation in KD-enhanced FL, mainly from the perspectives of black-box models of agents and differential privacy. 

\textbf{Black-box Models of agents.}
In vanilla FL, agents share their local model parameters with the server for global model aggregation, e.g., \texttt{FedAvg} \cite{mcmahan2017communication},  \texttt{FedDF} \cite{lin2020ensemble},  \texttt{FedAUX} \cite{sattler2021fedaux}, \texttt{FedBKD} \cite{lee2020edge}, \texttt{FedZKT}  \cite{zhang2021fedzkt}, \texttt{FedFTG} \cite{9879661}, \texttt{FedGen} \cite{zhu2021data}, etc. However, the model architecture and parameters are also sensitive information, as they can reveal private data through techniques like model inversion attacks \cite{fredrikson2015model,he2019model,wang2021variational}. In response, black-box models in collaborative learning have emerged as a privacy-preserving approach, where agents avoid sharing their local model parameters during collaboration \cite{pmlr-v97-hoang19a,chen2020online}. This approach enhances security by limiting exposure to potential attacks while still enabling effective knowledge exchange.

Considering black-box models, various global aggregation methods have been proposed, as shown in Table \ref{tab:related-work-FKD}. Most works, such as \texttt{FedMD} \cite{li2019fedmd}, \texttt{DS-FL} \cite{itahara2021distillation}, \texttt{Cronus} \cite{hongyan2019cronus}, \texttt{FD} \cite{jeong2018communication}, \texttt{HFD} \cite{8904164}, \texttt{FedProto} \cite{tan2021fedproto}, and \texttt{FedHKD} \cite{chen2023best}, utilize \textit{output averaging} for global aggregation, where agents share their local model outputs with the server. This method enables effective collaboration while maintaining the privacy of the agents' model parameters.

In addition to output averaging, \texttt{FCCL+} \cite{huang2023generalizable} incorporates \textit{intermediate feature sharing and averaging} for global aggregation, enabling more effective knowledge transfer. \texttt{FedAL} \cite{10606337} takes this further by training a \textit{discriminator} on the server to differentiate agents' local model outputs and promote consensus in agent behavior. \texttt{FedGD} \cite{zhang2023towards} trains a unified \textit{generator} on the server to produce public data, facilitating KD across agents.

Other approaches, including \texttt{CFD} \cite{sattler2020communication}, \texttt{MHAT} \cite{hu2021mhat}, and \texttt{FedGems} \cite{cheng2021fedgems}, train a \textit{global model} on the server using predictions provided by agents on a public dataset. This global model then generates predictions to facilitate KD on the client side. Additionally, \texttt{FedGKT} \cite{he2020group} and \texttt{FedDKC} \cite{wu2024exploring} train a \textit{global classifier} on the server using intermediate features and corresponding ground-truth labels shared by agents, enabling effective global knowledge aggregation.

In addition, \texttt{ADCOL} \cite{pmlr-v202-li23j} trains a \textit{discriminator} on the server to differentiate between agents' local model outputs, enabling the generation of consistent intermediate features across agents. \texttt{FedCKT} \cite{cho2021personalized} supports black-box models with varying architectures across agents, introducing a \textit{clustering} method to achieve personalized FL.

\textbf{Differential Privacy.}
In collaborative learning, agents share their local model parameters, such as gradients, with a central server to safeguard the privacy of their raw data. However, even this approach can inadvertently expose sensitive information, as sharing gradients among multiple agents may still lead to the privacy of raw data \cite{zhu2019deep}. 
Malicious agents can compromise collaborative learning systems through a variety of sophisticated attacks, aiming to degrade model performance, manipulate predictions, or extract sensitive user information. For instance, white-box attacks \cite{sun2020federated}, in which the adversary has full access to the model parameters and architecture, can include model inversion attacks, which attempt to reconstruct private training data from shared gradients or outputs. In addition, Byzantine agents \cite{qi2023differentially} may carry out integrity attacks by submitting malicious or manipulated model updates, thereby poisoning the global model and undermining its reliability.
In stealthy attack \cite{gong2022preserving}, adversaries craft subtle, undetectable behaviors that evade standard anomaly detection or robust aggregation mechanisms. These attacks often proceed slowly over multiple training rounds or mimic benign patterns to remain hidden while achieving malicious goals, such as implanting backdoors or introducing targeted errors.
Another serious threat is data reconstruction attacks \cite{xiao2024privacy}, where adversaries exploit intermediate representations or gradient information to infer or regenerate input data. Such attacks pose significant privacy risks, especially in collaborative learning environments where direct access to raw data is restricted.

Differential privacy (DP) is a robust privacy protection technique designed to ensure that individual data points within a dataset remain confidential. It achieves this by introducing statistical noise to the data or the queries made on it, thereby obscuring the contribution of any single data point. In the context of collaborative learning, differential privacy enables the construction of machine learning models that effectively utilize data from diverse sources while preserving the confidentiality of the underlying datasets. This approach strikes a critical balance between ensuring privacy and maintaining the utility of the learned model, allowing for collaborative insights without compromising individual data security.

Differential privacy in collaborative learning can be categorized into concentrated differential privacy (CDP) and local differential privacy (LDP). CDP, such as Zero-Concentrated Differential Privacy (zCDP) \cite{bun2016concentrated}, involves a central aggregator that applies noise to the aggregated model before any information are released. This approach is particularly valuable in FL, where there may be an adversary capable of accessing the global model and inferring whether a specific agent participated in the training process.
Conversely, LDP \cite{wei2021low} addresses the scenario of a curious but honest server, focusing on preserving the privacy of each agent's privacy in a local context. In this framework, each agent adds noise to their local models before sharing them with the central server. LDP is especially beneficial in sensitive applications, such as health records or personal messages. 

Traditional FL often employs LDP to protect agents' local model parameters, as demonstrated in works such as \cite{li2019asynchronous, sun2022profit, wei2020federated, kim2021federated, wei2021low, wei2021user}. In the context of \texttt{FedKT} \cite{li2020practical}, multiple agents engage in one-shot FL with a focus on protecting the privacy of both third-party participants and the server. KD has also been explored as a mechanism for model privacy protection. For instance, Shejwalkar et al. \cite{shejwalkar2019reconciling} utilized KD to develop a privacy-preserving model capable of withstanding membership inference attacks by training a protected model based on an original model.

In addition, certain studies have specifically emphasized differential privacy within the KD process. For example, \texttt{Rona} \cite{wang2019private} introduced privacy guarantees to compress large models for deployment on mobile devices by adding Gaussian noise to intermediate features and model outputs used for student model training. Similarly, Gao et al. \cite{gao2020private} aimed to secure the privacy of distillation loss between a pre-trained teacher network and a student model, employing adversarial learning to aid the student in effectively mimicking the teacher. These approaches highlight the potential of KD to balance model utility and privacy, making it an effective tool for secure model sharing and deployment in privacy-sensitive applications.

In the literature, research on KD-enhanced FL mainly centers around securing the LDP of agents, as shown in Table \ref{tab:related-work-dp}. This is typically achieved by protecting the privacy of agents' local model outputs over a public dataset, with methods such as adding Gaussian or Laplace noise to model predictions before transmission to the server. Examples include \texttt{FedMD-NFDP} \cite{sun2020federated}, \texttt{FedPDD} \cite{wan2023fedpdd}, \texttt{PrivateKT} \cite{qi2023differentially}, and \cite{gong2022preserving}.
On the server side, \texttt{FedMD-NFDP} and \texttt{FedPDD} employ an approach where the server averages the LDP-protected predictions and returns them to agents, maintaining privacy while enabling knowledge transfer among agents. In \texttt{PrivateKT}, the server fine-tunes the global model based on LDP-perturbed local model predictions and then distributes the updated global model to agents for the next round of model training. Meanwhile, \cite{gong2022preserving} proposes a one-shot ensemble distillation on the server, where knowledge is transferred from agents' LDP-protected outputs to a unified global model, thus integrating collective insights while upholding privacy standards.

Beyond protecting local model outputs, additional privacy-preserving strategies in KD-enhanced FL target other sensitive knowledge held by agents. For instance, \texttt{FedAUX} \cite{sattler2021fedaux} applies DP directly to agents' local model parameters, ensuring that privacy is preserved when training a global model on the server. Similarly, \texttt{FedAUXfdp} \cite{hoech2022fedauxfdp} perturbs agents' local model parameters using DP, allowing them to safely share these parameters with the server. The server initializes training based on the averaged perturbed models and refines the global model by utilizing a public unlabeled dataset, enhanced by agent-predicted labels. This approach strengthens model robustness against non-IID data while maintaining data privacy among agents.

Other methods, such as \texttt{FedHKD} \cite{chen2023best}, extend privacy measures to intermediate model features. DP is applied to intermediate representations before these are shared with the server, ensuring added layers of data confidentiality. Additionally, \cite{xiao2024privacy} addresses potential privacy risks by simulating an adversarial attack using generative adversarial networks (GANs) to estimate gradients and infer sensitive agent data from reported local model outputs on a public dataset. In response, they propose \texttt{LabelAvg}, which restricts information leakage by sharing only predicted hard labels, thus concealing inter-class similarity information of agents’ data.


\begin{table*}[t] \centering
    \caption{Related works on differential privacy in KD-empowered federated learning. }
    \label{tab:related-work-dp}
        \begin{center}
            \begin{small}
  \begin{tabular}{p{2.3cm}p{2.5cm}p{0.8cm}p{2.5cm}p{2.5cm}p{2.5cm}p{2.5cm}}
\hline 
Ref. & Attacks & DP & Agent to server & Server to agent & Data for knowledge extraction & Protected knowledge \tabularnewline
\hline
\texttt{FedMD-NFDP} \cite{sun2020federated} & white-box attacks & LDP & perturbed local model predictions &  average predictions & public dataset & model outputs \tabularnewline
\rowcolor{gray!15}\texttt{FedPDD} \cite{wan2023fedpdd} & white-box attacks & LDP & perturbed local model predictions &  average predictions & public dataset & model outputs \tabularnewline
\texttt{PrivateKT} \cite{qi2023differentially} & attacks by Byzantine agents & LDP & perturbed local model predictions &  global model & public dataset & model outputs \tabularnewline
\rowcolor{gray!15}
\cite{gong2022preserving} &stealthy attacks & LDP & perturbed local model predictions &  - (one-shot) & public dataset & model outputs \tabularnewline
\texttt{FedAUX} \cite{sattler2021fedaux}&- & LDP & local model parameters & global model parameter & private dataset & model parameters\tabularnewline
\rowcolor{gray!15}
\texttt{FedAUXfdp} \cite{hoech2022fedauxfdp}&white box attacks \& membership inference attacks& LDP & perturbed model parameter & feature extractor & public dataset & model parameters\tabularnewline
\texttt{FedHKD} \cite{chen2023best}& differential attacks&LDP & perturbed local intermediate features &  average intermediate features & private dataset & model intermediate features \tabularnewline
\rowcolor{gray!15}
\texttt{LabelAvg} \cite{xiao2024privacy}&data reconstruction attack \& white-box’ attacks & LDP & predicted hard label &  average predictions & public dataset & model outputs  \tabularnewline
\hline 
	\end{tabular}
\end{small}
\end{center}
\end{table*}

\subsection{Multi-Agent Domain Adaptation}
Domain adaptation \cite{farahani2021brief} focuses on transferring knowledge from a source domain, where labeled data is available, to a target domain with limited or no labeled data, often to improve model generalization across different but related distributions. 

There are various types of domain shifts recognized in the literature, with general domain shift representing a blend of these types \cite{kouw2018introduction}:
\begin{itemize} 
\item Data shift: This occurs when each class shares the same posterior distribution across domains, but the class distributions differ. An example of this shift type is fault detection settings. A common approach toward data shift involves using samples drawn from the source distribution and applying a re-weighting function to balance each class in alignment with the target distribution.
\item Covariate shift: In covariate shift, the posterior distribution is consistent across domains, but the data distribution itself varies. Scenarios like sample selection bias or missing data often experience covariate shifts. For covariate shifts, a correction function can be applied to source sample probabilities to reflect the target distribution accurately.
\item Concept shift: In this case, the data distribution remains constant, but the posterior distribution varies between domains. Examples include cases such as age-related changes in medical prognosis or non-stationary environments. Using labeled data from both domains, the posterior distributions can be measured to account for the shift.
\end{itemize}

In this subsection, we review the related works on domain adaptation across multiple agents from multi-source single-target domains and single-source multi-target domains respectively,


\textbf{Multi-source single-target domains.}
Various strategies address domain adaptation from multiple source domains to a single target domain.
Works in the literature mainly minimize the domain discrepancy through feature alignment and adversarial methods.

For feature alignment-based domain adaptation, \texttt{STEM}  \cite{Nguyen_2021_ICCV} comprehensively considers source domain diversity and data shift between the source and target domains to design an ensemble expert. Knowledge from this ensemble expert is distilled to the target model using KD.
In \texttt{MLAN} \cite{Xu_2022_WACV}, a joint alignment model is trained to learn domain-invariant features for both source and target domains, while separate models are trained for each target domain to capture target-specific knowledge. Mutual learning is applied between the joint alignment model and the fused target-specific models, enabling collaborative learning between the two branches.
Additionally, the authors in \cite{tong2021mathematical} propose a mathematical framework to quantify transferability in multi-source transfer learning. They linearly combine models learned from different source domains to adapt to the target domain and introduce an alternating iterative algorithm for training deep neural networks in multi-source transfer scenarios.

Some approaches label the target domain to support feature alignment. For instance, \texttt{KD3A} \cite{feng2021kd3a} uses models trained on multiple source domains to generate labels for unlabeled target data, then trains an extended model using this target data with predicted labels. KD is employed to transfer knowledge from the source domains to the extended model. By combining the source domain models and the extended model, \texttt{KD3A} achieves improved performance on the target domain.
In \texttt{UAD} \cite{song2024multi}, two levels of regularization are proposed for multi-source domain adaptation. At the model level, domain gaps between each source model and the target domain are minimized through confidence-guided model distillation. At the instance level, inference confidence is calculated for each data sample, with aggregated predictions from source models used to label target data. These aggregated labels are then applied to train the target model.
In \texttt{CoNMix} \cite{Kumar_2023_WACV}, the target model is trained using pre-trained source models without access to labeled source data. To support this, pseudo data and labels are generated by taking a convex combination of two randomly sampled data points. This strategy facilitates KD from multiple source models to the target model, helping to prevent overfitting.

For adversarial learning-based domain adaptation, \texttt{FADA} \cite{peng2019federated} allows each agent to access both its source domain data and target domain data. Each agent trains a separate feature extractor for its source and target domains, while adversarial learning is applied on the server to distinguish between source and target domains. This setup promotes knowledge transfer from multiple source domains to the target domain. \texttt{FADA} also optimizes aggregation weights to combine predictions from source domains, improving final prediction accuracy.
Similarly, \texttt{MDDA} \cite{zhao2020multi} maps the target domain into the representation space of pre-trained models from multiple source domains by employing a min-max game to align representations effectively.
In \texttt{SSD} \cite{10029939}, pseudo-labels for target domain data are generated using trained models from multiple source domains, with domain importance valuations. A domain discriminator is built based on the distilled source domains to identify the dominant source domain and capture relationships between source and target domains. This approach fine-tunes both source and target models, enhancing the reliability of target domain labeling.

 \textbf{Single-source multi-target domains.}
The aim of single-source multi-target domain adaptation is to train a model that effectively operates across multiple target domains, utilizing a model that has been pre-trained on a single source domain.
In \texttt{KD-MTDA} \cite{belal2021knowledge} and \texttt{MT-MTDA} \cite{Nguyen-Meidine_2021_WACV}, the source domain data is utilized to train models that adapt to each target domain through unsupervised learning. The models for the target domains serve as teacher models to train a common student model that applies to all target domains via KD.
In \texttt{CCL} \cite{Isobe_2021_CVPR}, the authors propose simultaneously training a expert model for each target domain to imitate the outputs of the model trained from the source domain. They also implement regularization techniques to bring different experts closer together. Ultimately, online knowledge distillation from multiple experts is applied to train a unified target model.
\texttt{UDA} \cite{Saporta_2021_ICCV} introduces a multi-discriminator framework, where each discriminator is trained to distinguish between the source and target domains, as well as to differentiate a specific target domain from other target domains. 

\subsection{Federated Multi-Modal Learning}
In this subsection, we review the literature on federated multi-modal learning from two primary objectives: modality fusion and modality expansion, as summarized in Table \ref{tab:related-work-multi-modal}. For a comprehensive overview of multi-modal federated learning, refer to \cite{lin2023federated,che2023multimodal,thrasher2023multimodal}; notably, \cite{thrasher2023multimodal} focuses on applications in healthcare.

\textbf{Modality fusion.} Modality fusion indicates combining information from multiple data modalities to improve the overall model’s performance. In healthcare, for instance, combining patient data from medical images, lab reports, and health records can lead to better diagnostic predictions than relying on a single data type.

Related work on modality fusion emphasizes the transfer of knowledge from agents to the server with diverse data modalities. For instance, in \cite{wang2020multimodal}, models trained on each data modality serve as teachers to train a student model on the server that operates with incomplete modalities. Recognizing that application data often features incomplete modalities, \cite{9567675} trains models using multi-modal data and distills the knowledge from these models to a server model that utilizes uni-modal data.

In a federated setting, the server aims to aggregate the local models from agents to create a global model that accommodates all modalities. For example, \texttt{FedCola} \cite{sun2023exploring} facilitates collaboration among uni-modal agents with different data modalities. On the server, a multi-modal global model is trained, incorporating both modality-specific and modality-shared parameters. Aggregation occurs only between agents with the same modality. Considering the partial participation of agents, the participating modalities in each round are balanced. 
\texttt{FedDAT} \cite{chen2024feddat} focuses on fine-tuning a foundation model for agent-specific multi-modal data, aggregating local adapter models on the server. Mutual knowledge distillation is employed between the global and local models to mitigate overfitting.
\texttt{PT-FUCH} \cite{li2023prototype} trains a global prototype for each modality on the server by clustering the features of local models from agents. This approach distills knowledge from the global prototype to local models, helping to avoid overfitting and inaccuracies in local semantic representation.
\texttt{edCMD} \cite{bano2024fedcmd} trains a teacher model using labeled data from one modality and transfers knowledge to a student model with unlabeled data via KD. Parameter averaging is utilized on the server to aggregate the student models from agents.
To address the challenges posed by heterogeneous model architectures among agents and the server, \texttt{CreamFL} \cite{u2023multimodal} allows agents with limited computational resources to maintain smaller models than the server. A public multi-modal dataset is accessible on both the server and all agents, with agents uploading their local model outputs to the server for aggregation. The server then provides feedback on the global model outputs to facilitate local KD.

\textbf{Modality expansion.} Modality expansion aims to extend an existing model to include additional modalities without retraining it from scratch, thus broadening the model’s scope to incorporate new data types or adapt to new tasks. For example,  an AI model initially trained to analyze textual data in electronic health records could be expanded to also process imaging data, enabling it to make more holistic patient assessments without a complete retraining.
\texttt{MKE} \cite{Xue_2021_ICCV} aims to adapt a pre-trained uni-modal model to perform the same task using additional, unlabeled multi-modal data. Cross-modal distillation is employed to transfer knowledge from the uni-modal model to a multi-modal model by learning from the extra modality. Results demonstrate that the multi-modal student model generalizes better than its uni-modal teacher.

\begin{table*}[t] \centering
    \caption{Related works on multi-modal FL enhanced by KD. }
    \label{tab:related-work-multi-modal}
        \begin{center}
            \begin{small}
  \begin{tabular}{p{2.3cm}p{2.8cm}p{3cm}p{3cm}p{3cm}}
\hline 
Ref. & Scenario & Agent & Server & Shared knowledge \tabularnewline
\hline
\cite{wang2020multimodal} & modality fusion & uni-modality & incomplete modality & model parameters \tabularnewline
\rowcolor{gray!15}
\cite{9567675} & modality fusion & multi-modality & uni-modality & model outputs\tabularnewline
\texttt{FedCola} \cite{sun2023exploring} & modality fusion & uni-modality & - & model parameters 
\tabularnewline
\rowcolor{gray!15}
\texttt{FedDAT} \cite{chen2024feddat} & modality fusion & multi-modality & - & model parameters\tabularnewline
\texttt{PT-FUCH} \cite{li2023prototype} & modality fusion & multi-modality & - & model fused feature
\tabularnewline
\rowcolor{gray!15}
\texttt{FedCMD} \cite{bano2024fedcmd} & modality fusion & multi-modality & - & model parameters \tabularnewline
\texttt{CreamFL} \cite{u2023multimodal} &  modality fusion & multi-modality & public multi-modality & model representations 
\tabularnewline
\rowcolor{gray!15} 
\texttt{MKE} \cite{Xue_2021_ICCV} & modality expansion & uni-modality & another uni-modality & model parameters\tabularnewline
\hline 
	\end{tabular}
\end{small}
\end{center}
\end{table*}

\subsection{Federated Continual Learning}
Surveys on continual learning and related methods can be found in \cite{qu2021recent, wang2024comprehensive, 10341211}. Most continual learning research focuses on the less-forgetting of old tasks while simultaneously learning new ones.
For example, \texttt{LwF} \cite{li2017learning} constrains model updates by limiting the divergence between the new dataset and previous ones through KD. Building on \texttt{LwF}, \texttt{LwM} \cite{dhar2019learning} incorporates an additional attention distillation loss to regularize changes in attention on the new task relative to the previous one.
\texttt{GedDL}  \cite{Christian2021MGeoCont} proposes distilling the feature space of the previous model into the current one by considering the cosine similarity between projections along the geodesic flow on a low-dimensional manifold.
\texttt{iCaRL} \cite{8100070} addresses a class-incremental scenario, where each base learner trains a model to include a new class in the dataset. An example dataset is maintained to facilitate knowledge transfer from the previous class to the new one.
\texttt{EEIL} \cite{castro2018end} presents an end-to-end incremental learning framework that adapts the model to a new class of data. The outputs of previous models are used to distill their knowledge into the new model.

In a federated setting, a comprehensive survey of regularization methods for transfer learning in domain-incremental FL and an experimental comparison can be found in \cite{huang2023survey}. For a survey specifically on FCL, refer to \cite{yang2024federated}, compared to which our paper primarily focuses on KD-empowered FCL approaches.
We summarize the less-forgetting strategies in Table \ref{tab:related-work-less-forgetting}, categorizing them based on scenario, loss function for less forgetting, operating data, and the model used for computing the loss function. The scenarios indicate the specific tasks associated with FCL, which include general FL, class-incremental FL, domain-incremental FL, and task-incremental FL, as described below:
\begin{itemize} 
\item Class-incremental scenario: This involves training models to incorporate new classes of data while retaining knowledge of previously learned classes. 
\item Domain-incremental scenario: Here, the focus is on adapting models to new domains od data while preserving performance on older domains. 
\item Task-incremental scenario: This scenario addresses the challenge of learning multiple tasks sequentially, ensuring that the model maintains performance across all tasks. \end{itemize}
In this subsection, we review KD-empowered FCL approaches from various perspectives, including model recalling, replay buffer, data generation, prompting techniques, and selective forgetting.

\textbf{Model recalling.} Model recalling refers to the practice of utilizing historical models to represent the  knowledge from previous datasets, thereby achieving less forgetting. For instance, \cite{dong2022federated} preserves earlier models and minimizes the distance between the new and previous models through KD.

\textbf{Replay buffer.} 
Some works maintain a replay buffer of old data to facilitate knowledge extraction and replay. For example, the authors in \cite{Fedrecon2021} extend \texttt{iCaRL} \cite{8100070} to federated learning settings, alleviating forgetting when learning incremental classes by utilizing the stored old dataset.
\texttt{FL-IIDS} \cite{jin2024fl} dynamically maintains an example dataset to help agents avoid forgetting previous tasks. To achieve this, KD is applied on agents using the augmented local dataset with example samples. Additionally, a class gradient balance loss is constructed to regulate the learning rate for new knowledge and the forgetting rate for old knowledge, ensuring that the best old model is selected for KD.
In \texttt{FedPMR} \cite{wang2023federated}, a replay buffer retains a limited number of data samples from earlier tasks to support less forgetting. A parameter decomposition mechanism is employed to better extract generic and agent-specific parameters.
\texttt{CFeD} \cite{ma2022continual} utilizes a public dataset to mitigate intra-task forgetting in FCL through KD on both the agent and server sides. On the agent side, the KL divergence between the agents' local model outputs and the outputs of the global model from previous tasks is minimized. On the server side, the agents' model outputs serve as knowledge to be distilled into the global model.
In \texttt{DAG} \cite{carta2023projected}, each agent maintains a small set of out-of-distribution data. Less forgetting is achieved through KD between the intermediate features of the current local model and the global model from the previous task.
\texttt{FedET} \cite{liu2023fedet} employs a Transformer model to preserve the local knowledge of agents. Each agent contributes a small number of local data samples to facilitate KD from previous tasks to new tasks.
\texttt{Re-Fed} \cite{li2024towards} selects and stores model-relevant data during training, using a mix of new data and replayed samples for model updates.
\texttt{LGA} \cite{dong2023no} considers three types of agents in class-incremental FL: agents with both old and new categories of data, agents with only old categories, and agents with only new categories. To address category imbalance among agents, the authors propose a category-balanced gradient-adaptive compensation loss. They also introduce a category gradient-induced semantic distillation loss to mitigate local catastrophic forgetting, minimizing the KL divergence between the outputs of old and new global models for each class pair.

\textbf{Data generation.} When maintaining a replay buffer on agents is not feasible due to privacy concerns and high memory costs, agents can generate data samples to support less-forgetting regularization. \texttt{TRINA} \cite{li2024facing} employs a diffusion model to generate data that facilitates knowledge transfer from previous tasks to the new task model. Additionally, a novel weight computation method is proposed to achieve balanced global model aggregation on the server.
\texttt{TARGET} \cite{zhang2023target} and \texttt{MFCL} \cite{babakniya2024data} train a generator using the global model obtained from the last task to produce synthetic data. This synthetic data is then utilized for KD between the local model and the previous global model.

\textbf{Prompting techniques.}
Instead of maintaining a memory-costly replay buffer, \texttt{FCILPT} \cite{liu2023federated} encodes both task-relevant and task-irrelevant knowledge into prompts. The method minimizes the cross-entropy loss of the new data over the most relevant prompts, thereby alleviating catastrophic forgetting of old classes.

 \textbf{Selective forgetting.}
 While learning new tasks, it may be necessary to forget certain knowledge to protect data privacy. In \cite{shibata2021learning}, the authors investigate the process of forgetting specific classes when training on a new dataset. They propose a mnemonic code to represent the associated information of each class. For a new task, a synthetic image is generated for each class using the mnemonic code, creating an augmented dataset to train the model for the new task. By selecting the retained mnemonic codes, the method enables control over which classes are remembered or forgotten.
\texttt{AF-FCL} \cite{wuerkaixi2023accurate} emphasizes the drawbacks of retaining biased or irrelevant features, which can arise in federated scenarios due to statistical heterogeneity. The authors introduce the concept of accurate forgetting, employing KD in the feature space to mitigate complete forgetting. They approximate the true feature distribution for each class as a multivariate Gaussian and generate global data features. These features are selected for model training based on their correlation with the current task.

\subsection{Federated Multi-Task Learning}
Research on federated multi-task learning with KD primarily focuses on enhancing knowledge transfer across different tasks while addressing the unique challenges posed by distributed data.
For instance, in \texttt{UM-Adapt} \cite{Kundu_2019_ICCV}, models for various tasks are sequentially trained by distilling knowledge from previous tasks, enabling continuous learning and adaptation.
In \texttt{FedLinked} \cite{kadar2022fedlinked}, the authors evaluate cross-utility between pairs of tasks to facilitate representation regularization using a public dataset. This method enhances the collaborative learning process by ensuring that representations are aligned across tasks.
\texttt{FedICT} \cite{wu2023fedict} takes a novel approach by leveraging the assumed prior knowledge of agents' local data distributions, thereby eliminating the need for a public dataset. The authors introduce a similarity-based local knowledge adjustment loss function to effectively distill global knowledge to the local side, with similarity computed based on the local data distributions of the agents.
\texttt{FedEM} \cite{marfoq2021federated} approaches the problem by considering the dataset of each task as a mixture of underlying distributions. The weights assigned to each task are evaluated and utilized for iterative cross-task knowledge transfer through an Expectation-Maximization algorithm during model training. This approach allows for a more nuanced understanding of the relationships between tasks, improving overall model performance.

\subsection{Federated Knowledge Graph Embedding}
KD has proven effective for KGE \cite{ge2024knowledge} in various applications, including KGE model compression and knowledge transfer within federated KGE frameworks, as summarized in Table \ref{tab:related-work-kge}.

For KGE compression, KD enables the distillation of essential knowledge from large, complex KGE models into smaller, more efficient ones, reducing computational overhead while preserving critical relational information. 
In \texttt{G-CRD} \cite{joshi2022representation}, KD is applied to transfer knowledge from a high-dimensional pre-trained Graph Neural Network (GNN) teacher to a lower-dimensional student model. The approach leverages contrastive learning to closely align the representations of the student model with those of the teacher, ensuring that the student effectively captures complex relational information. \texttt{IterDE} \cite{liu2023iterde} adopts an iterative distillation approach, training a low-dimensional student knowledge graph (KG) from a high-dimensional teacher. At each iteration, a progressively smaller student is trained from the current student model (or directly from the teacher in the initial step), promoting incremental compression without significant performance loss.
\texttt{PMD} \cite{fan2024progressive} enhances the KD process by applying mask-based representations on both student and teacher models, facilitating the capture of rich semantic features for distillation. The model applies progressive distillation, transferring information from different parts of the teacher’s representation space to the student. Similarly focusing on KGE compression, \texttt{DulDE} \cite{zhu2022dualde} strengthens conventional KD by assigning adaptive weights to the hard loss and the KD loss, optimizing the trade-off between preserving intricate details and compressing the model size. 
 
For federated KGE \cite{liu2024federated}, KD  facilitates knowledge transfer across distributed agents, enhancing model performance on decentralized data while maintaining data privacy. This dual utility of KD highlights its adaptability and effectiveness in addressing both scalability and data security challenges in KGE. 
\texttt{FedE} \cite{chen2021fede} pioneered federated learning in KGE, addressing data privacy and heterogeneity in decentralized KGE training. Expanding on \texttt{FedE}, Zhang et al. \cite{zhang2024low} introduced an adaptive asymmetric temperature scaling mechanism. This technique allows agents with low-dimensional KGE models to collaboratively learn from a high-dimensional, pre-trained teacher model, enhancing cross-agent knowledge transfer while managing dimensional discrepancies.
To improve knowledge exchange between server and agents, \texttt{FedLU} \cite{zhu2023heterogeneous} applies mutual KD, where local embeddings on the agents and global embeddings on the server continuously transfer knowledge back and forth. This bidirectional KD mechanism promotes efficient global knowledge absorption while retaining local specifics. In \texttt{FedSpray} \cite{fu2024federated}, a global feature-structure encoder is trained on the server to produce soft targets that act as guidance for regularizing local GNN models via KD. 

Furthermore, KD plays an integral role in advancing continual and multi-teacher KGE. In \texttt{IncDE} \cite{liu2024towards}, KD addresses the continual KGE challenge by employing incremental distillation, which mitigates catastrophic forgetting as the model sequentially learns new entities and relationships over rounds. This method enables the model to retain knowledge from previous training rounds, preserving valuable information about earlier entities.
For multi-teacher KGE, \texttt{MulDE} \cite{wang2021mulde} leverages knowledge from multiple pre-trained teachers. In this framework, a junior student model is trained to aggregate and integrate knowledge from diverse teacher models. Additionally, a senior student model serves as a mediator, selecting and refining the most informative entities from each teacher by identifying entities with top-K scores. This two-level student design ensures that the junior model receives distilled knowledge that is both broad and contextually relevant, enhancing its ability to generalize across different knowledge sources.

\begin{table*}[t] \centering
    \caption{Related works on KGE via KD. }
    \label{tab:related-work-kge}
        \begin{center}
            \begin{small}
 \begin{tabular}{p{2.5cm}p{3cm}p{5cm}p{3.8cm}}
\hline 
Ref. &  Objective & Approach & Shared knowledge \tabularnewline
\hline
\texttt{G-CRD} \cite{joshi2022representation} & KGE compression & contrastive learning & - \tabularnewline
\rowcolor{gray!15}\texttt{IterDE} \cite{liu2023iterde}  &  KGE compression & iterative training &-\tabularnewline
\texttt{PMD} \cite{fan2024progressive}&   KGE compression &progressive distillation & - \tabularnewline
\rowcolor{gray!15}
\texttt{DulDE} \cite{zhu2022dualde} &   KGE compression & weight adjustment & - \tabularnewline
\texttt{FedE} \cite{chen2021fede}  & federated KGE & parameter averaging & local/global embedding models \tabularnewline
\rowcolor{gray!15}
\cite{zhang2024low}   & federated KGE & adaptive asymmetric temperature scaling & local/global embedding models\tabularnewline
\texttt{FedLU} \cite{zhu2023heterogeneous} & federated KGE & mutual KD & local/global embedding models\tabularnewline
\rowcolor{gray!15}
\texttt{FedSpray} \cite{fu2024federated} & federated KGE & feature-structure encoding & local structure proxy\tabularnewline
\texttt{IncDE} \cite{liu2024towards}  & continual KGE & incremental distillation & - \tabularnewline
\rowcolor{gray!15}
\texttt{MulDE} \cite{wang2021mulde} & multiple-teacher KGE & senior\& junior students & - \tabularnewline
\hline
	\end{tabular}
\end{small}
\end{center}
\end{table*}

\section{KD empowered Memory Consolidation under Hierarchical Collaborative Pattern}

This section reviews the application of KD for memory consolidation within hierarchical collaborative frameworks. 
In hierarchical collaborative learning, aggregating all agents on a central server can lead to significant synchronization delays and increased vulnerability to attacks. To address these challenges, edge servers can be utilized to establish a hierarchical collaborative structure.

In this framework, each edge server maintains a regional model for the agents it covers. The training process proceeds through the following stages:
\begin{itemize}
    \item Local Training: Each agent downloads its local model from the corresponding edge server and trains it using its local data.
    \item Edge-Level Aggregation: Each edge server aggregates the locally updated models from the agents under its coverage. 
    \item Cloud-Level Aggregation: The central server then aggregates the regional models uploaded from the edge servers. This layer of aggregation helps to unify the knowledge acquired across different regions, enhancing the overall model performance while minimizing communication overhead.
\end{itemize}
It is important to note that during edge-level aggregation, agents may conduct multiple iterations of local training, while cloud-level aggregation can involve several edge-level aggregations. This hierarchical approach not only improves efficiency but also enhances the robustness of the collaborative learning system.

Most existing work considers vanilla FL with parameter averaging. For work involving KD, existing research primarily focuses on optimizing agent clustering, addressing non-IID data distributions among large-scale agents, and accommodating heterogeneous model architectures. For instance, \cite{nguyen2024hierarchical} introduced a hierarchical federated learning framework that employs KD to facilitate model aggregation between a central server and multiple edge servers, each responsible for aggregating model parameters from local agents.
To tackle the challenges posed by non-IID data and the forgetting phenomenon in hierarchical FL, the authors of \cite{nguyen2024hierarchical} proposed global and regional regularization techniques using KD to stabilize local training processes. This approach helps mitigate the impact of data variability and enhances model retention over time.
In \cite{lackinger2024inference}, the authors focused on optimizing the locations of local servers and their allocation to end agents to minimize total resource costs. Model parameters are shared among agents, edge servers, and the central server for model aggregation. This work considers the heterogeneous resource capacities of local servers and the varying inference request rates of agents, ensuring efficient resource management in a hierarchical structure.
Additionally, \texttt{FedHKT} \cite{deng2023hierarchical} emphasizes the importance of data heterogeneity among agents in federated learning. The authors designed a hierarchical FL framework that accommodates agents with heterogeneous model architectures, considering homogeneous intra-cluster data distributions and heterogeneous inter-cluster distributions. They proposed methods for agent-edge and edge-cloud knowledge transfer, facilitating more effective collaboration and knowledge sharing across diverse agents. Model parameters are exchanged between agents and the edge server for parameter averaging, while model outputs are exchanged between edge servers and the central server to enable global knowledge transfer.

\section{KD empowered Memory Consolidation under Decentralized Collaborative Pattern}
The workflow in decentralized collaborative learning generally follows these steps:
\begin{itemize}
    \item Local training: Each agent independently trains its local model using its dataset and shares its learned knowledge, such as model parameters or outputs, with neighboring agents.
    \item Decentralized model aggregation: Each agent then aggregates the knowledge received from its neighbors. The aggregated model is subsequently shared with neighbors, facilitating a collaborative knowledge transfer or continuing to the next training iteration.
\end{itemize}

Research on decentralized learning has primarily aimed at refining aggregation techniques, such as adjusting agent weights, and optimizing communication topologies among agents.
For instance, \texttt{D-FedEM} \cite{marfoq2021federated} addresses decentralized multi-task learning by allowing each agent to compute weights for underlying data distributions to transfer knowledge across tasks. Each agent computes these weights based on the parameters shared by neighboring models, enabling task-specific learning in a decentralized structure.
Other studies explore the role of network topology on decentralized learning performance. Vogels et al. \cite{vogels2023beyond} investigate topological structures like ring, fully connected, and star formations, demonstrating that convergence rates and stability vary with each topology, influencing the overall learning process. Further work by Xing et al. \cite{xing2021federated} and Koloskova et al. \cite{koloskova2020unified} has analyzed convergence error when the communication topologies change unpredictably.
TCPL \cite{xu2021decentralized} investigated how to construct optimal topologies in decentralized federated learning networks with time-varying connectivity. 

In Table \ref{tab:related-work-decentralized}, we summarize the approaches in decentralized knowledge distillation, where all of them assume fixed communication topologies. Some studies focus on framework design with static weights for model aggregation, while others aim to optimize agent-specific aggregation weights.
For instance, \texttt{D-Distillation} \cite{bistritz2020distributed} allows agents to communicate in a sparse, fixed peer-to-peer topology. Each agent gathers information from its neighbors and distills their knowledge. The framework achieves convergence with probability 1.
\texttt{IDKD} \cite{ravikumar2023homogenizing} employs high-confidence public data for KD across agents, facilitating knowledge exchange in decentralized environments without direct sharing of private data. In \texttt{DFML} \cite{khalil2024dfml}, one agent serves as a central aggregator. It collects model parameters from other participants and aggregates them by performing KD over its local dataset. This updated model, serving as a the new global model, is then broadcast back to the agents.

On the other hand, some approaches prioritize adaptive weight optimization. \texttt{DLAD} \cite{ma2021adaptive} enhances aggregation performance by adjusting weights based on each agent’s local confidence. A discriminator evaluates data reliability, assigning higher weights to agents with more confident predictions. Jeong et al. \cite{jeong2023personalized} introduced a decentralized FL framework where a randomly selected agent acts as a star node, aggregating its neighbors’ models. The aggregation weights are determined based on the statistical distance between model parameters.

\begin{table*}[t] \centering
    \caption{Related works on the decentralized pattern. }
    \label{tab:related-work-decentralized}
        \begin{center}
            \begin{small}
  \begin{tabular}{p{2.8cm}p{2.8cm}p{3cm}p{2.3cm}p{3cm}}
\hline 
Ref. & Changing topology? & Weight determination & public dataset & Shared knowledge \tabularnewline
\hline
\texttt{D-Distillation}\cite{bistritz2020distributed} & \xmark & fixed & \cmark & model outputs\tabularnewline
\rowcolor{gray!15}
\texttt{IDKD} \cite{ravikumar2023homogenizing} & \xmark & fixed & \cmark & model outputs \tabularnewline
\texttt{DFML} \cite{khalil2024dfml} & \xmark & fixed & \xmark & model parameters\tabularnewline
\rowcolor{gray!15}
\texttt{DLAD} \cite{ma2021adaptive} & \xmark & confidence based on discriminator' output & \cmark & model outputs \tabularnewline
\cite{jeong2023personalized} & \xmark & based on statistic distance & \xmark & model parameters\tabularnewline
\hline

	\end{tabular}
\end{small}
\end{center}
\end{table*}

\section{Challenges \& Future Directions}
In this section, we analyze the remaining challenges associated with utilizing memory mechanisms and KD in collaborative learning and propose potential future directions to address these challenges.

\subsection{Tradeoff Between Memory and Knowledge}
In collaborative learning across multiple agents, knowledge can be shared in two forms: abstract and concrete. Abstract knowledge is readily accessible and can be directly retrieved from procedural memory. In contrast, concrete knowledge requires extraction through model inference, a process that necessitates the use of shared data and additional computational resources across the agents. Meanwhile, concrete knowledge helps maintain the privacy of each agent’s local model architecture and parameters. Therefore, it is crucial to strike a balance between memory usage, knowledge sharing, computational costs, and the protection of model privacy. A hybrid knowledge sharing mechanism could be an effective solution to address the diverse resource capacities and privacy requirements of agents.

\subsection{Evaluation Metrics}
Current work commonly evaluates the effectiveness of KD using posterior metrics such as loss functions and model accuracy, which reveal KD’s impact only after the process has been conducted. This reactive approach means that there is limited foresight into KD’s effectiveness before training begins. Some existing studies have identified factors that influence KD efficiency. For instance, \texttt{TAKD} \cite{mirzadeh2020improved} highlights how variations in teacher and student model architectures can significantly affect KD outcomes. However, there remains a lack of understanding regarding the influence of multi-dimensional heterogeneity between teacher and student models, such as the effects of data heterogeneity, architectural differences, and their hybrid impacts on KD efficacy. Addressing this challenge requires the development of proactive evaluation metrics that can predict KD performance prior to training, offering a clearer roadmap for successful KD integration in collaborative learning.

 
\subsection{Resource-constrained Collaborative KD}
Existing research addresses the computation and communication resource constraints of agent devices primarily by utilizing techniques from traditional federated learning, such as model pruning and quantization \cite{9355797}. However, with the integration of KD, there are additional promising strategies that can be explored to adapt more flexibly to the limitations and heterogeneity of resources. 
 
One promising topic is to incorporate the partial participation of agents in collaborative KD. 
Allowing agents to partially participate in each training round can further alleviate the computational burden on individual agents during collaborative KD. Currently, research has primarily focused on partial participation designs within conventional collaborative learning frameworks \cite{10621105}. In the context of collaborative KD, it is essential to account for the unique characteristics of KD, particularly regarding computation and communication overheads, to effectively and adaptively implement partial participation among agents. 

Additionally, another effective strategy for reducing the computational overhead in collaborative learning is split federated learning \cite{10631047,thapa2022splitfed}, which offloads part of the local model training to the server, thereby alleviating the computational burden on agents. Current research in split federated learning has primarily concentrated on homogeneous model architectures and fixed split positions among agents. However, exploring split federated KD with heterogeneous model architectures and flexible split positions presents a promising avenue for enhancing collaborative KD. This approach could facilitate more efficient knowledge transfer while accommodating the diverse capabilities of participating agents.

Furthermore, KD holds significant promise for scenarios involving heterogeneous end devices, particularly in fine-tuning large foundation models for specific downstream tasks on user devices. This approach enables both model compression and personalization, as knowledge can be distilled from the foundation model to the local models of agents. It is essential to design effective collaborative frameworks among agents and develop KD mechanisms. Thus, we can improve the adaptability and performance of models across diverse user devices, as well as the scalability of collaborative learning systems, ultimately leading to more tailored solutions for specific applications.

\subsection{Memory Management for Collaborative KD}
In real-world deployments, particularly on memory-constrained edge devices, effective memory management is crucial for enabling scalable and efficient collaborative learning. KD offers a promising avenue for optimizing memory usage in such settings \cite{li2024flexnn}.

One key challenge arises in layer-wise streaming inference, where the peak memory usage is typically dominated by the layer with the highest memory demand. In this context, KD can be leveraged to redesign or simplify the neural network architecture, encouraging a more balanced memory profile across layers and mitigating peak memory constraints during inference.

Moreover, reducing memory consumption in deep neural network (DNN) inference often incurs additional computational and I/O overhead beyond standard inference procedures. For instance, enabling layer-wise streaming requires frequent loading of layer weights from external memory, which introduces considerable memory I/O operations. This overhead is particularly pronounced in systems with Unified Memory Architecture (UMA) and limited storage bandwidth, which is common in edge and mobile devices, where the memory access latency becomes a significant bottleneck.
Therefore, selecting appropriate KD strategies for collaborative learning and inference, i.e., output sharing or parameter sharing strategies, is essential not only to preserve model performance but also to minimize overall memory overhead.

\subsection{Incentive Mechanism}
Agents, such as organizations, often exhibit self-interested behavior, seeking to absorb knowledge from others while minimizing the disclosure of their own local knowledge. In addition to model performance, these agents prioritize resource costs, privacy, and fairness within the collaborative environment. In extreme cases, this self-interest can lead to a complete refusal among agents to share their local knowledge, undermining the potential for effective collaboration. While existing research has explored incentive mechanisms within the context of traditional federated learning \cite{9565851}, these approaches do not consider the specific dynamics introduced by knowledge distillation (KD). Therefore, it is crucial to develop tailored incentive mechanisms that specifically address the characteristics of collaborative KD, stressing agents' preference for high model performance, low resource cost, high privacy, and fairness.

\subsection{Decentralized KD with Adaptive Topology}
Current research on decentralized KD mainly focuses on designing decentralized collaboration frameworks, as shown in Table \ref{tab:related-work-decentralized}. However, further investigation is needed to optimize communication topologies, which play a critical role in enhancing the communication efficiency and model performance of decentralized KD. This requires accurately evaluating KD effectiveness, which is challenging because the local model parameters of individual clients are often unknown after involving output-sharing KD. 
Moreover, such designs should accommodate dynamic and heterogeneous conditions, such as varying bandwidths and time-dependent connectivity between peers, to support effective collaborative KD across diverse and evolving environments.

\subsection{Theoretical analysis}
Existing studies have analyzed the performance of KD-enhanced collaborative learning, as outlined in Table \ref{tab:related-work-FKD}. The theoretical analysis of these works includes generalization bounds, primarily based on learnability and VC Dimension theories \cite{ben2010theory,ben2006analysis}, of the obtained models and the convergence rates of model training. However, only a few studies have concentrated on generalization bounds \cite{huang2023generalizable,cho2021personalized,10606337,lin2020ensemble,pmlr-v202-li23j,zhu2021data,wang2023dafkd} and convergence rates \cite{cho2021personalized,10606337,chen2023best,he2022learning,ozkara2021quped}.
Moreover, current findings on generalization bounds and convergence rates do not fully account for KD efficiency across varying conditions, such as data and model heterogeneity. Furthermore, theoretical insights specific to diverse collaborative learning tasks remain underexplored. Addressing these gaps is essential for advancing theoretical analysis in KD across different scenarios. This would not only refine understanding but also support practical system optimization, including strategies for resource-constrained settings, incentive mechanism development, and decentralized communication topology optimization, etc.

\section{Conclusion}\label{sec:conclusion}
In this paper, we have explored the role of memory and knowledge in knowledge distillation within collaborative learning, highlighting their importance in addressing challenges such as model heterogeneity, data diversity, resource disparities, and privacy concerns across various collaborative tasks. We define and categorize the types of memory and knowledge involved to provide a clearer understanding of how KD facilitates knowledge transfer among heterogeneous agents. Furthermore, we discuss the remaining challenges and future directions, focusing on improving evaluation methods, real-world deployment under resource constraints, the impact of self-interested agents, communication network topologies, and enhancing interpretability in large-scale collaborative learning systems.


%




\ifCLASSOPTIONcaptionsoff
  \newpage
\fi



%
\bibliographystyle{IEEEtran}
\bibliography{ref}
\end{document}